\documentclass[12pt,a4paper,final]{iopart}
\newcommand{\bea}{\begin{eqnarray}}
\newcommand{\eea}{\end{eqnarray}}
\newcommand{\be}{\begin{equation}}
\newcommand{\ee}{\end{equation}}

\newcommand{\ra}{\rangle}
\usepackage{iopams}
\usepackage{iopams}
\usepackage{graphicx}
\usepackage[breaklinks=true,colorlinks=true,linkcolor=blue,urlcolor=blue,citecolor=blue]{hyperref}
\begin{document}
\title{Radiation properties of an oscillating atom in the presence of external fields}

\author{F. Chegini$^{1}$}
\ead{chegini.sepideh@gmail.com}

\author[cor]{F. Kheirandish$^{1}$}
\eads{\mailto{f.kheirandish@uok.ac.ir}}

\author{M. R. Setare$^{1}$}
\ead{rezakord@ipm.ir}
\vspace{1cm}
\address{$^1$Department of Physics, Faculty of Science, University of Kurdistan, P.O.Box 66177-15175, Sanandaj, Iran}

\begin{abstract}
In the first part of the present work, the correction to photon emission rate of an oscillating two-level atom in the presence of electromagnetic quantum vacuum field has been investigated for two different configurations: (i) Atom is trapped in the vicinity of a perfect conductor (ii) Atom is trapped between two perfect conductors. In the second part, the correction to the decay rate of an initially excited oscillating two-level atom due to the presence of a perfect conducting surface is found.
\end{abstract}

\pacs{
42.50.Lc,	
31.30.jf,	
31.15.-p,	
42.50.Ct,	
}
\vspace{2pc}
\noindent{\it Keywords}: Photon emission rate, Two-level atom, Quantum vacuum, Decay rate, Oscillating atom

\submitto{}
%
%
\section{Introduction}

The quantum vacuum fluctuations is a ubiquitous result of quantizing classical field theories. One of the most important classical field theories with extensive applications in science and technology is the Maxwell's electromagnetic field theory. The electromagnetic quantum vacuum fluctuations is a direct result of quantizing the electromagnetic field theory that its manifestation can be approved from experimental observations like Casimir effect \cite{1,2,3,c1,c2,c3,c4,c5,c6,c7} Lamb shift of atomic transitions \cite{l1,l2,l3,l4} and spontaneous emission of an initially excited atom \cite{l4,4,5,6,7,8,9}. In addition to quantum vacuum fluctuations, there may be some external fields. Then the fluctuating field satisfies predefined boundary conditions at the location of external fields. For example, in the case of electromagnetic vacuum fluctuations, the presence of metallic or dielectric materials leads to a nonhomogeneous electromagnetic energy density due to boundary conditions causing Casimir forces among the material fields \cite{r1,k}. Also, the spectroscopic measurements of atoms shows that the transition frequencies of an atom in the presence of boundary conditions differ from their free space value \cite{10}. Therefore, the radiation properties of atoms will change when they are placed in front of a metallic or dielectric surface \cite{l3,l4}. On other hand, the dynamical interaction between light and moving atoms in the realm of non-relativistic atom optics has been studied extensively in related fields like laser cooling and trapping and quantum optomechanics \cite{ors,opto}. A neutral atom can interact with the electric component of the electromagnetic field through its electric dipole moment, also a moving electric dipole carries a magnetic dipole moment that can interact with the magnetic component of the electromagnetic field. Therefore, a moving atom can couple to both electric and magnetic components of the electromagnetic field. Consequently, the moving atoms or equivalently dipoles, exhibit the phenomena like R\"{o}ntgen quantum phase shift \cite{Aha-Bohm, Wilkens}. Barton and Calegoracos investigated the spontaneous emission of atoms moving in a classically assigned trajectory \cite{Barton}. Also, Muller studied the role of the vacuum fluctuations in spontaneous excitation of a uniformly accelerated atom in its ground state \cite{Muller}. This process corresponds to the famous Unruh effect \cite{Unruh}. In general, Unruh effect expresses that the vacuum state is different for inertial and accelerated observers. The vacuum state for the inertial observer means no particle state but the accelerated observer detects a thermal bath of particles at temperature $T$. In \cite{11}, the nonrelativistic oscillation of an atom in its ground state has been investigated in free space. In the present article, we have generalized the problem studied in \cite{l3,11} to the case where there are boundary conditions due to the presence of metallic surfaces that modify the Green tensor of the fluctuating field. In the absence of the boundary conditions the results are in agreement with those reported in \cite{l3,11}.

The main motivations for the present work are: (i) In order to have more control on the photon emission rate of an oscillating two-level atom, some boundary conditions like the presence of conductors have been considered as a generalization of the results reported in \cite{11} (ii) From the experimental point of view holding an atom in an exact distance from a surface is impossible due to the Heisenberg uncertainty relations, the best we can do is trapping the atom in an average distance from a surface. Therefore, the decay rate of an initially excited two-level atom needs to be corrected due to the oscillation of the atomic center of mass.

The paper is organized as follows: In Sec. II, the basic formulation is presented and correction to photon emission rate of an oscillating two-level atom in the presence of electromagnetic quantum vacuum field has been investigated in two different cases: (i) Atom is trapped in vicinity of a perfect conductor (ii) Atom is trapped between two perfect conductor. In Sec. III, the spontaneous decay rate of an initially excited atom oscillating in the vicinity of a perfect conductor has been investigated. Finally, we conclude in Sec.IV.

\section{The basic formalism}

In this section we introduce very briefly the basic material and notations that will be used in the following subsections. Let us consider a two-level atom described by the ground state $|g\ra$ and excited state $|e\ra$ with energies $E_g$ and $E_e$ respectively. The internal frequency is defined by $(E_e-E_g)/\hbar=\omega_0$ and let $\mathbf{r}(t)$ denotes the center of mass motion of the moving atom with corresponding velocity $\mathbf{v}(t)=\dot{\mathbf{r}}(t)$. Then the interaction Hamiltonian in dipole approximation in CGS units can be written as
\begin{eqnarray}\label{1}
&& H_{int}=-\mathbf{d}.\left[\textbf{E}(r(t))+\frac{\textbf{v}(t)}{c}\times\textbf{B}(r(t))\right],
\end{eqnarray}
where electric and magnetic field operators are in interaction picture and are evaluated at the position of the atom. Let the atom be initially in its ground state, then the single-photon emission rate can be calculated from the time-dependent perturbation theory as
\begin{eqnarray}\label{2}
&& \Gamma = \frac{1}{T} \frac{1}{\hbar^2} \left|\int_0 ^T{\left\langle 1_{\mathbf{q}, \lambda},e\left|H_{int}(t)\right|0,g \right\rangle}dt\right|^2,
\end{eqnarray}
where $T$ is the duration of the interaction. In Eq. (\ref{2}) by $|0,g\ra$ we mean that the atom is in its ground state $|g\ra$ and the field state is the vacuum state $|0\ra$. Similarly, $|e,1_{\mathbf{q}, \lambda}\ra$ means that the atom is in its excited state and the field state is a single-particle state described by a photon with wave number $\mathbf{q}$ and polarization $\lambda$.

The electromagnetic field operators can be decomposed into positive $\textbf{E}^+\,(\textbf{B}^+)$ and negative $\textbf{E}^-\,(\textbf{B}^-)$ frequency parts as
\begin{eqnarray}\label{3}
&& \textbf{E}(\textbf{r},t)=\frac{1}{\sqrt{2\pi}}\int_0 ^{\infty}{d\omega \left[\textbf{E}^+(\textbf{r},\omega)e^{-i\omega t}+\textbf{E}^-(\textbf{r},\omega)e^{i\omega t}\right]},\nonumber\\
&& \textbf{B}(\textbf{r},t)=\frac{1}{\sqrt{2\pi}}\int_0 ^{\infty}{d\omega \left[\textbf{B}^+(\textbf{r},\omega)e^{-i\omega t}+\textbf{B}^-(\textbf{r},\omega)e^{i\omega t}\right]}.
\end{eqnarray}
The dipolar matrix elements in interaction picture can be written as $\hat{d}_i(t)=\hat{d}_i(0) e^{-i \omega_0 t}$. For a spherically symmetric atomic state the matrix elements of the electric dipole satisfy
\begin{eqnarray}\label{4}
&& \overline{\langle e | \hat{d}_i(0) |g \rangle  \langle g |\hat{d}_j(0) |e \rangle}=\delta_{i j} | \langle e | \textbf{d}(0) |g \rangle |^2/3.
\end{eqnarray}
Using Eqs. (\ref{1})-(\ref{4}), we find that the single-photon emission rate has contributions from quadratic terms in electric and magnetic fields given by
\begin{eqnarray}\label{5}
&& \Gamma=\Gamma^{EB}+\Gamma^{EE}+\Gamma^{BB},
\end{eqnarray}
where
\begin{eqnarray}\label{6}
\Gamma^{EB} &=& \frac{1}{T} \frac{ | \langle e | \mathbf{d}(0) |g \rangle |^2}{6 \pi{\hbar}^2}\int_{0}^{\infty}{d \, \omega} \int_{0}^{\infty}{d \, \omega '}\delta_{i l} \int_{0}^{T}{d t} \int_{0}^{T}{d t'} e^{-i \, \omega_0 (t-t')} e^{-i\omega t+i \omega' t'} \nonumber\\
&& \times \, \bigg[ \langle  0 | \hat{E}^+ _i(\mathbf{r}(t),\omega) \Big(\frac{ \mathbf{v}(t')}{c}\times \hat{ \mathbf{B}}^-(\mathbf{r}(t'),\omega') \Big)_l |0  \rangle\nonumber\\
&& +\langle 0 | \,  \Big(\frac{\mathbf{v}(t)}{c}\times \hat{ \mathbf{B}}^+(\mathbf{r}(t),\omega) \Big)_i \hat{E}^- _l(\mathbf{r}(t'),\omega') | 0 \rangle \bigg],
\end{eqnarray}
\begin{eqnarray}\label{7}
\Gamma^{EE} &=& \frac{1}{T} \frac{ | \langle e | \mathbf{d}(0) |g \rangle |^2}{6 \pi{\hbar}^2}\int_{0}^{\infty}{d \, \omega} \int_{0}^{\infty}{d \, \omega '}\delta_{i l} \int_{0}^{T}{d t} \int_{0}^{T}{d t'} e^{-i \omega_0(t-t')} e^{-i \omega t+i \omega' t'} \nonumber\\
&& \times\langle 0 | \hat{E}^+ _i(\mathbf{r}(t),\omega)  \hat{E}^- _l(\mathbf{r}(t'),\omega')| 0 \rangle,
\end{eqnarray}
\begin{eqnarray}\label{8}
\Gamma^{BB} &=& \frac{1}{T} \frac{| \langle e | \mathbf{d}(0) |g \rangle |^2}{6 \pi{\hbar}^2}\int_{0}^{\infty}{d \, \omega} \int_{0}^{\infty}{d \, \omega '}\delta_{i l} \int_{0}^{T}{d t} \int_{0}^{T}{d t'} e^{-i \, \omega_0 (t-t')} e^{-i\omega t+i \omega' t'} \nonumber\\
&& \times \, \bigg[ \langle  0 | \Big(\frac{ \mathbf{v}(t)}{c}\times \hat{ \mathbf{B}}^+(\mathbf{r}(t),\omega) \Big)_i \, \Big(\frac{\mathbf{v}(t')}{c}\times( \hat{ \mathbf{B}}^-(\mathbf{r}(t'),\omega') \Big)_l | 0 \rangle \bigg],
\end{eqnarray}
and the repeated subscript indices ($i,j,l=x,y,z$) are summed over. Eq. (\ref{6}) can be rewritten as
\begin{eqnarray}\label{9}
\Gamma^{EB} &=& -\frac{1}{T} \frac{ | \langle e | \mathbf{d}(0) |g \rangle |^2}{6 \pi{\hbar}^2  c}\int_{0}^{\infty}{d \, \omega} \int_{0}^{\infty}{d \, \omega '}  \int_{0}^{T}{d t} \int_{0}^{T}{d t'} e^{-i \, \omega_0 (t-t')} e^{-i\omega t+i \omega' t'} \nonumber\\
&& \times \, \bigg[ \langle  0 | \hat{A}^+ _i(\mathbf{r}(t),\omega) \hat{A}^- _i(\mathbf{r}(t'),\omega')  |0 \rangle  \bigg(\omega' \ (\frac{\mathbf{v}(t)}{c} \cdot \mathbf{q} )+\omega \ (\frac{\mathbf{v}(t')}{c} \cdot \mathbf{q}' ) \bigg) \bigg],\nonumber\\
\end{eqnarray}
where we made use of the following identities
\begin{eqnarray}\label{10}
&&\hat{E}^+ _i(\mathbf{r},\omega)=(i \omega/c) \hat{A}^+ _i(\mathbf{r},\omega),\quad \hat{\mathbf{B}}^+=\nabla \times \hat{\mathbf{A}}^+ ,\nonumber\\
&&\hat{E}^- _i(\mathbf{r},\omega)=(-i \omega/c) \hat{A}^- _i(\mathbf{r},\omega),\quad \hat{\mathbf{B}}^-=\nabla \times \hat{\mathbf{A}}^- .
\end{eqnarray}
The fluctuation-dissipation relation connects the correlation functions of the vector potential components to the imaginary part of the electromagnetic Green tensor as \cite{14}
\begin{eqnarray}\label{11}
&& \langle 0 |\hat{A}^+ _i(\textbf{r},\omega)\,\hat{A}^- _j(\textbf{r}',\omega')|0 \rangle=2{\hbar}\  \mbox{Im}[\ G_{ij}]\, (\textbf{r},\textbf{r}',\omega)\ \delta (\omega-\omega').
\end{eqnarray}
The electromagnetic Green tensor in a medium described by the dielectric function $\varepsilon(\mathbf{r},\omega)$ satisfies the equation
\begin{eqnarray}\label{12}
&& \bigg(q^2\,\varepsilon(\mathbf{r},\omega)\,\delta_{mi}-\frac{\partial^2}{\partial x_m\,\partial x_i}+\delta_{mi}\ \nabla^2\bigg)\ G_{ij}\ (\textbf{r},\textbf{r}',\omega)= - {4 \pi }\delta_{mj}\ \delta(\textbf{r}-\textbf{r}'),\nonumber\\
\end{eqnarray}
where $q^2=\omega^2/c^2$. Therefore, the contribution $\Gamma^{EB}$ can be rewritten in terms of the Green tensor as
\begin{eqnarray}\label{13}
  \Gamma^{EB} &=& \frac{1}{T} \frac{| \langle e | \mathbf{d}(0) |g \rangle |^2}{3 \pi{\hbar c}}\int_{0}^{\infty}{d \, \omega} \, \omega \int_{0}^{T}{d t} \int_{0}^{T}{d t'} e^{-i(\omega+\omega_0)(t-t')}\nonumber\\
  && \times \Big[-\frac{\mathbf{v}(t)}{c} \cdot \mathbf{q} -\frac{\mathbf{v}(t')}{c} \cdot \mathbf{q} \Big]
 \,\mbox{Im}[G_{i i} ( \mathbf{r}(t), \mathbf{r}(t'),\omega)].
\end{eqnarray}
Similarly, we have
\begin{eqnarray}\label{14}
\Gamma^{EE} &=& \frac{1}{T} \frac{| \langle e | \mathbf{d}(0)|g \rangle |^2}{6 \pi{\hbar}^2c^2}\int_{0}^{\infty}{d \, \omega} \, \omega  \int_{0}^{\infty}{d \, \omega '} \, \omega'  \int_{0}^{T}{d t} \int_{0}^{T}{d t'} e^{-i( \omega_0)(t-t')} \nonumber\\
&&\times \bigg[ e^{-i\omega t+i \omega' t'}
\,\langle 0 | \hat{A}^+_i (\mathbf{r}(t),\omega)  \hat{A}^-_i (\mathbf{r}(t'),\omega')| 0 \rangle \bigg],\nonumber\\
&=& \frac{1}{T} \frac{ | \langle e | \mathbf{d}(0) |g \rangle |^2}{3 \pi{\hbar}c^2} \int_{0}^{\infty}{d \, \omega} \, {\omega^2} \int_{0}^{T}{d t}\int_{0}^{T}{d t'} e^{-i(\omega+ \omega_0)(t-t')}\nonumber\\
&& \times\,  \mbox{Im}[G_{i i}(\mathbf{r}(t),\mathbf{r}(t'),\omega)].
\end{eqnarray}
\begin{eqnarray}\label{15}
 \Gamma^{BB} &=& \frac{1}{T} \frac{| \langle e | \mathbf{d}(0) |g \rangle |^2}{6 \pi{\hbar}^2}\int_{0}^{\infty}{d \, \omega} \int_{0}^{\infty}{d \, \omega '}\delta_{i l} \int_{0}^{T}{d t} \int_{0}^{T}{d t'} e^{-i \, \omega_0 (t-t')} \nonumber\\
&& \times\bigg\{ e^{-i\omega t+i \omega' t'}
\bigg( \langle  0 | \Big[\frac{ \mathbf{v}(t)}{c}\times \Big(\mathbf{q} \times \hat{ \mathbf{A}}^+(\mathbf{r}(t),\omega) \Big) \Big]_i \nonumber\\
&& \times\, \Big[\frac{\mathbf{v}(t')}{c}\times \Big( \mathbf{q}' \times \hat{ \mathbf{A}}^-(\mathbf{r}(t'),\omega')\Big) \Big]_l | 0 \rangle \bigg) \bigg\},\nonumber\\
&=& \frac{1}{T} \frac{| \langle e | \mathbf{d}(0) |g \rangle |^2}{3 \pi{\hbar} c^2} {\int_0^T{dt}}{\int_0^T{dt'}} {\int_0^{\infty}{d{\omega}}} \,{{e^{-i(\omega+\omega_0)(t-t')}}} \nonumber\\
&& \times\,[v(t)v(t')q^2\delta_{j k}-(\textbf{q}\times\textbf{v}(t))_{j}(\textbf{q}\times\textbf{v}(t'))_{k}]\,\mbox{Im}[G_{j k}(\mathbf{r}(t),\mathbf{r}(t'),\omega)].\nonumber\\
\end{eqnarray}
Recently, the population of one-photon states resulting from the non-relativistic oscillatory motion of an atom initially prepared in its ground state and interacting with quantum vacuum in free space, has been investigated in \cite{11}. In the following subsections we generalize this problem to the case where there are external material fields like the presence of perfect conductors.
\begin{figure}[htb!]
\centering
  \includegraphics[scale=0.30]{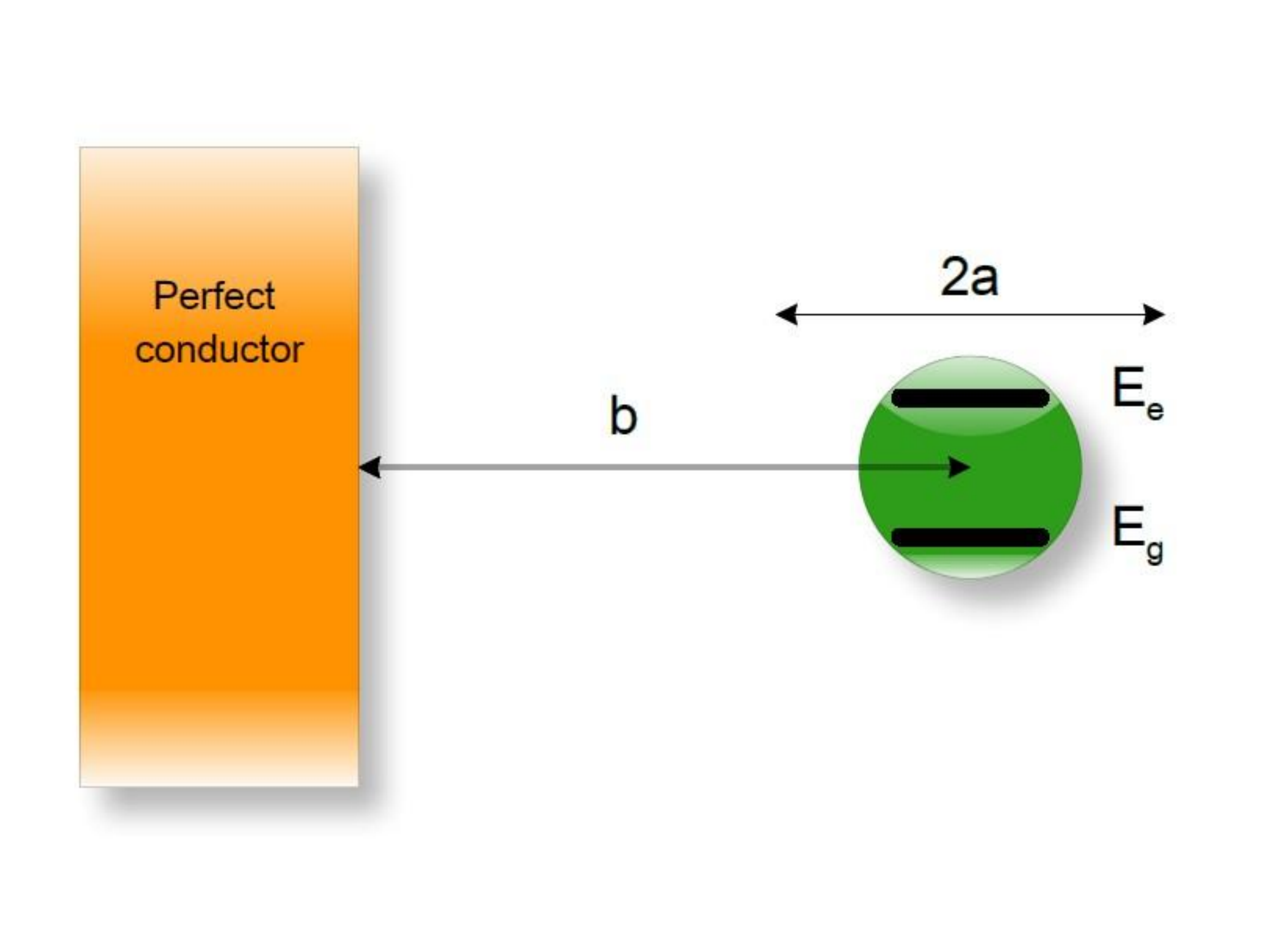}
  \caption{(Color online) A two-level atom oscillating in non-relativistic regime with frequency $\omega_{cm}$ in the vicinity of an ideal conducting plate. The parameter $b$ denotes the distance from the center of oscillation: $z(t)=b+a\,\cos(\omega_{cm} t),\,\,\,(b>a) $.}
\end{figure}
\subsection{Oscillating atom in the vicinity of a perfect conductor}

In this section we will find the single-photon emission rate from the non-relativistic oscillatory motion of a two-level atom in front of a perfect conductor. The atom moves along the $z$-axis which is perpendicular to the conductor surface. In this case the dielectric function is defined by
\be\label{16}
\varepsilon(z,\omega)=\left\{
                        \begin{array}{ll}
                          \infty, & z\leq 0 \\
                          1, & z> 0
                        \end{array}
                      \right.
\ee
In this case, the electromagnetic Green tensor can be obtained easily following the method applied in \cite{16} (see Appendix A).
The location of the trapped atom along the $z$-axis at time $t$ is given by $\textbf{r}(t)=z(t)\,\hat{k}=[b+a\cos{(\omega_{cm}t)}]\,\hat{k}$ ($\omega_{cm}$ is the frequency of oscillation) and implicitly we have assumed that $b> a$, see Fig.1. We have
\begin{eqnarray}\label{17}
&& \left|\textbf{r}(t)-\textbf{r}(t')\right|= \left|z(t)-z(t')\right|,\nonumber\\
&& z(t)+z(t')=2b+a [\cos{(\omega_{cm}t)}+\cos{(\omega_{cm}t'})], \nonumber\\
&& z(t)-z(t')=a[\cos{(\omega_{cm}t)}-\cos{(\omega_{cm}t')}].
\end{eqnarray}
After straightforward calculations, we find from Eqs. {\ref{13}}, (\ref{14}) and (\ref{15}) (Appendix A)
\begin{eqnarray}\label{18}
\Gamma^{EB} &=& \frac{1}{6} \Gamma_0 \frac{v_{max}^2}{c^2} \Big(\frac{\omega_{cm}}{\omega_0}-1\Big)^3 \Big(\frac{\omega_0}{\omega_{cm}}-1\Big),\nonumber\\
\Gamma^{EE} &=&  \frac{1}{4}  \Gamma_0 \frac{v_{max}^2}{c^2}\Big(\frac{\omega_{cm}}{\omega_0}-1\Big)^3\Big(1-\frac{\omega_0}{\omega_{cm}}\Big)^2 \nonumber\\
&&\times\,\bigg[ \frac{1}{3}+\sin{(2b \, q_0)} \Big(\frac{3}{4}\frac{1}{b^5 q_0^5}-\frac{3}{2}\frac{1}{b^3 q_0^3}+\frac{1}{2}\frac{1}{b \, q_0} \Big)\nonumber\\
&& +\cos{(2 b \, q_0)}\, \Big(\frac{-3}{2}\frac{1}{b^4 q_0^4}+\frac{1}{b^2 q_0^2} \, \Big) \bigg],\nonumber\\
\Gamma^{BB} &=&  \frac{1}{4} \Gamma _0 \frac{v_{max}^2}{c^2} \Big(\frac{\omega_{cm}}{\omega_0}-1\Big)^3 \bigg[ \frac{2}{3}+{\cos{(2b \, q_0)}} \Big(\frac{-3}{4}\frac{1}{b^2 q_0^2}+\frac{3}{4} \frac{1}{b^4 q_0^4} \Big)\nonumber\\
&& +{\sin{(2b \, q_0)}} \Big(\frac{7}{8}\frac{1}{b^3 q_0^3}-\frac{1}{2}\frac{1}{b \, q_0}-\frac{3}{8}\frac{1}{b^5 q_0^5} \, \Big) \bigg],
\end{eqnarray}
where we have assumed $0\leq q_0=(\omega_{cm}-\omega_0)/c$, $v_{max}= \omega_{cm}a$, and $\Gamma_0$ is the free space emission rate
\begin{eqnarray}\label{19}
&& \Gamma_0= \frac{4 \left|\left\langle e\left| \textbf{d}(0)\right|g\right\rangle\right|^2\,{\omega_0}^3}{3\hbar\,c^3}.
\end{eqnarray}
Therefore,
\begin{eqnarray}\label{20}
\Gamma &=& \Gamma_1 \bigg\{1+3 \, \frac{(1-\frac{\omega_0}{\omega_{cm}})^2}{1+(\frac{\omega_0}{\omega_{cm}})^2} \bigg[ \sin{(2b\,q_0)}\bigg(\frac{3}{4}\frac{1}{b^5q_0^5}-\frac{3}{2}\frac{1}{b^3q_0^3}+\frac{1}{2}\frac{1}{b\, q_0}\bigg)\nonumber\\
&& +\cos{(2b\,q_0)}\bigg(\frac{-3}{2}\frac{1}{b^4q_0^4}+\frac{1}{b^2q_0^2}\bigg)\bigg]\nonumber\\
&& +\frac{3}{1+(\frac{\omega_0}{\omega_{cm}})^2} \bigg[\sin{(2b\,q_0)}\bigg(\frac{-3}{8}\frac{1}{b^5q_0^5}+\frac{7}{8}\frac{1}{b^3q_0^3}+\frac{-1}{2}\frac{1}{b\, q_0}\bigg)\nonumber\\
&& +\cos{(2b\,q_0)}\bigg(\frac{3}{4}\frac{1}{b^4q_0^4}+\frac{-3}{4}\frac{1}{b^2q_0^2}\bigg)\bigg]\bigg\}.
\end{eqnarray}
\begin{figure}[htb!]
\includegraphics[width=18.0pc]{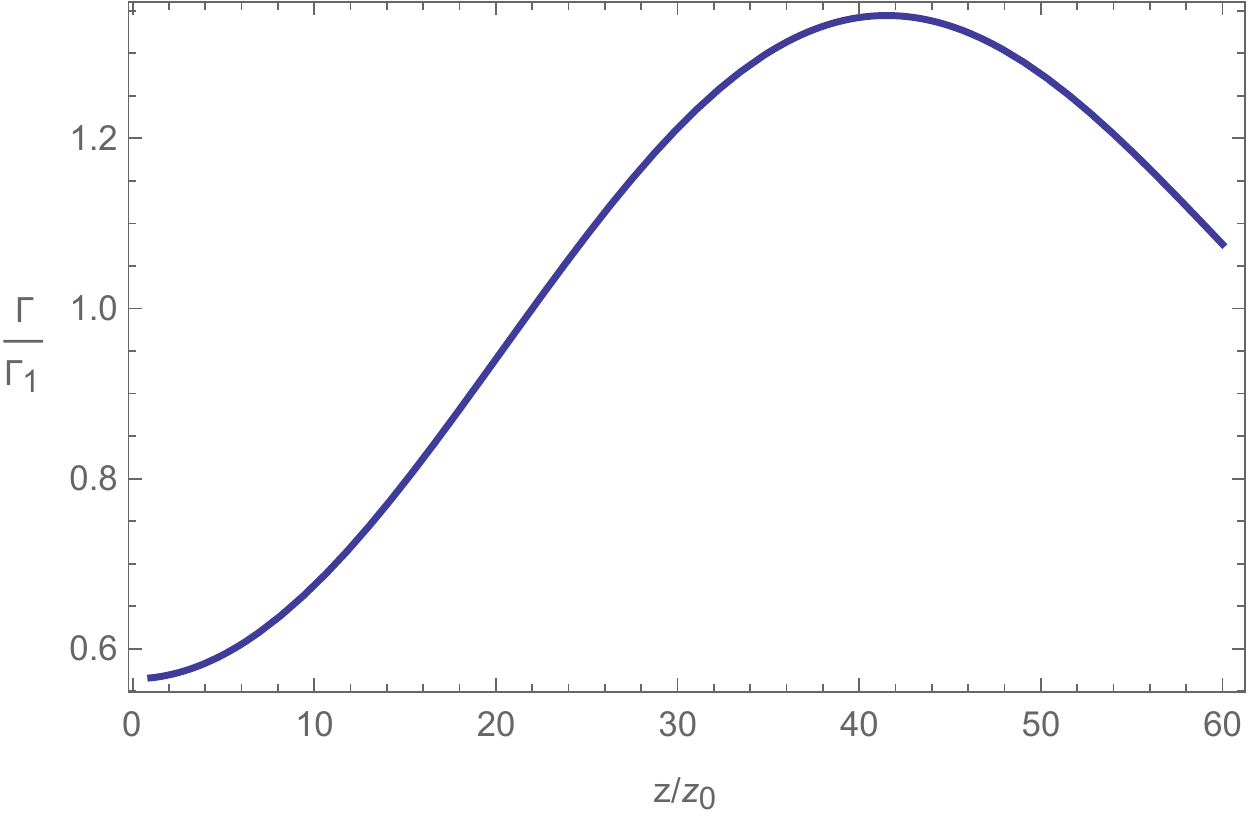}
\hspace{5mm}
\vspace{0.4cm}
\includegraphics[width=18.0pc]{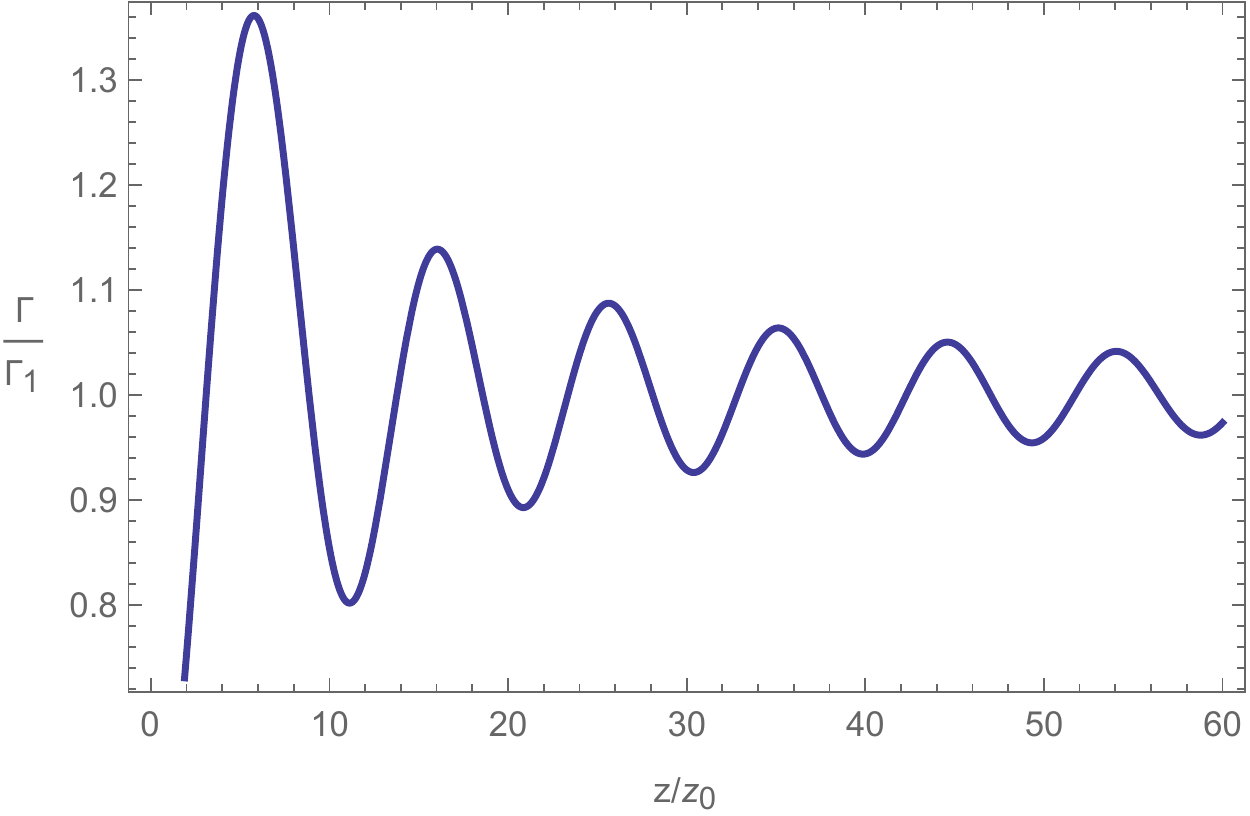}
\vspace{0.4cm}
\caption{(Color online) The scaled emission rate $\Gamma/\Gamma_1$ of a non-relativistic oscillating two-level atom located in front of a perfectly conducting plate in terms of the scaled distance ($z/z_0=2\, \omega_{cm} \, b/c$) for $\omega_{cm} = 1.1 \,\omega_0 $ ( The left plot ) and for $\omega_{cm} = 3\,\omega_0 $ (The right plot )}  \label{fig2}
\end{figure}
When the atom is far enough from the plate, the radiation rate tends to the radiation rate in free space $\Gamma_1$
\begin{eqnarray}\label{21}
&& \Gamma_1=\frac{1}{12}\,\Gamma _0 \frac{v_{max}^2}{c^2} \bigg(1+ \Big(\frac{\omega_0}{\omega_{cm}}\Big)^2 \bigg) \Big(\frac{\omega_{cm}}{\omega_0}-1\Big)^3,\,\,\,(\omega_{cm}\geq \omega_0),
\end{eqnarray}
that had been considered in \cite{11}. The result reported in \cite{11} slightly differs from Eq. (\ref{21}) may be due to typographical errors. In Fig. (2), the scaled emission rate $\Gamma/\Gamma_1$ of a non-relativistic oscillating two-level atom located in front of a perfectly conducting plate has been depicted in terms of the scaled distance ($z/z_0=2 \, \omega_{cm} \, b/c$) for $\omega_{cm} = 1.1 \, \omega_0$ and $\omega_{cm} = 3 \, \omega_0$.
\subsection{The emission rate of an oscillating atom between perfect parallel conductors }

In order to have more control on quantum dynamics of the oscillating two-level atom we consider the geometry depicted in Fig.3.
\begin{figure}[htb!]
\centering
  \includegraphics[scale=0.40]{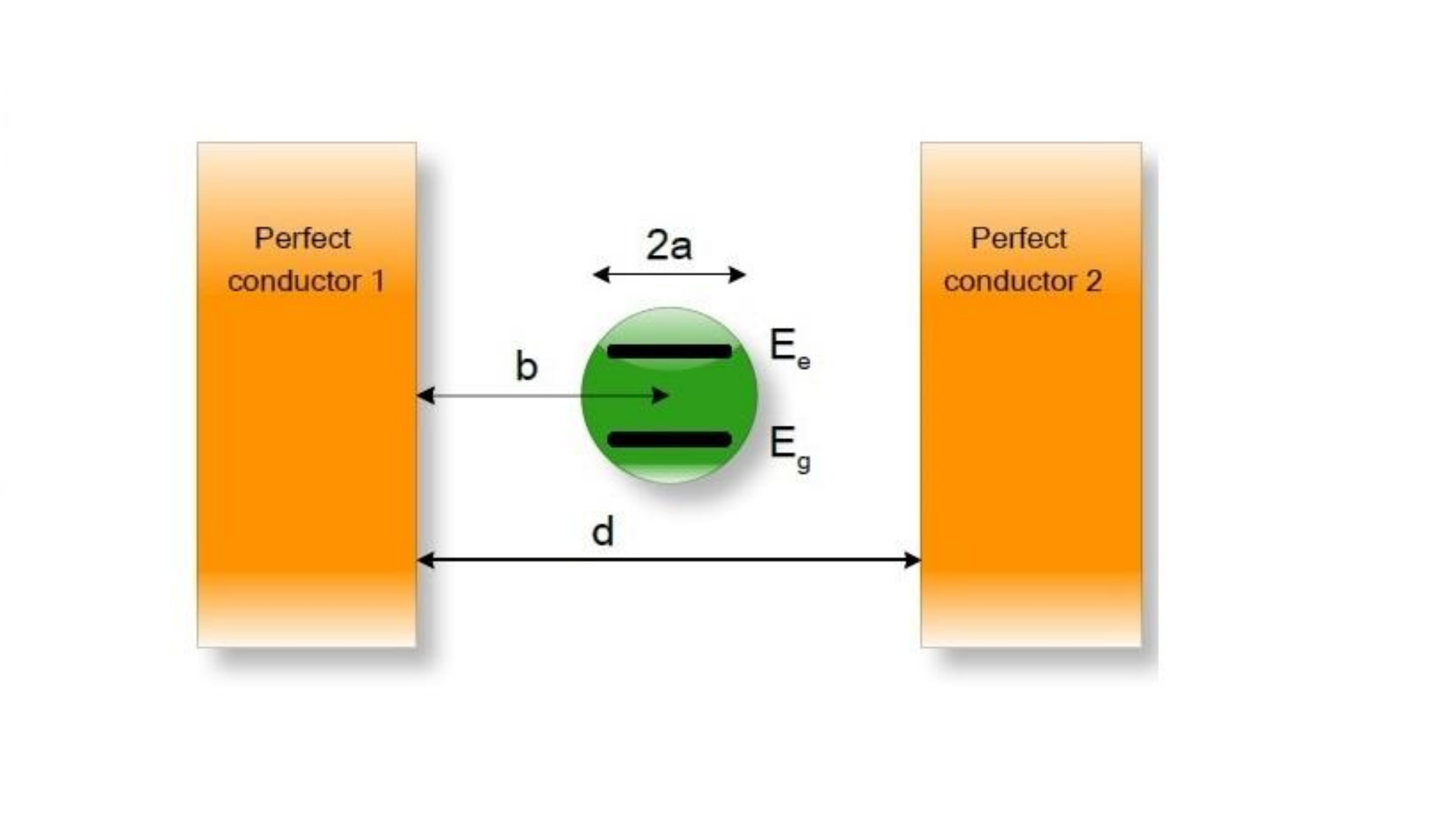}
  \vspace{0.5cm}
\caption{(Color online) A two-level atom oscillating in non-relativistic regime with frequency $\omega_{cm}$ between ideal conducting plates. The parameter $b$ denotes the distance from the center of oscillation: $z(t)=b+a\,\cos(\omega_{cm} t),\,\,\,(d>b>a) $.}
\end{figure}
\be\label{22}
\varepsilon(z,\omega)=\left\{
                        \begin{array}{ll}
                          \infty, & z\leq 0 \\
                         1, & d > z> 0 \\
												\infty, & z\geq d
                        \end{array}
                      \right.
\ee
Following calculations similar to the previous section (see Appendix B) we find
\begin{eqnarray}\label{23}
\Gamma^{EB} &=& \frac{1}{6} \Gamma_0 \frac{v_{max}^2}{c^2} \Big(\frac{\omega_{cm}}{\omega_0}-1\Big)^3 \Big(\frac{\omega_0}{\omega_{cm}}-1\Big),\nonumber\\
\Gamma^{EE} &=&  \frac{\left| \left\langle e\left| \mathbf{d}(0)\right|g\right\rangle \right|^2}{3{\hbar}} \,\frac{v_{max}^2}{c^2} \frac{(\omega_{cm}-\omega_0)^2}{\omega_{cm}^2} \nonumber\\
&& \times\,\mbox{Im} \bigg[ \int_0^{q_0} {d u}  \bigg(u^2 \cot{(u \, d)}+\frac{u^4}{q_0^2} \big(\cot{(u \, d)} \, {\cos{(2u \, b)}}+\sin{(2u \, b)} \big) \bigg) \bigg],\nonumber\\
\Gamma^{BB} &=&   \frac{\left| \left\langle e\left| \mathbf{d}(0)\right|g\right\rangle \right|^2}{6{\hbar}} \frac{v^2_{max}}{c^2}\mbox{Im} \bigg[ \int _0^{q_0}{d \, u} \, (q_0^2+u^2)\nonumber\\
&& \times\,\bigg(\cot{( u\, d)} -\frac{u^2}{q^2_0}(\cos{(2u\, b)} \, \cot{(u \, d)}+\sin{(2u \, b)} ) \bigg) \bigg],
\end{eqnarray}
where $u=\sqrt{q^2_0-q_{\parallel}^2}$. The emission rate is
\begin{eqnarray}\label{24}
\Gamma &=& \Gamma_1 \bigg\{1-\frac{2}{1+(\frac{\omega_0}{\omega_{cm}})^2}-\frac{(1-\frac{\omega_0}{\omega_{cm}})^2}{1+(\frac{\omega_0}{\omega_{cm}})^2} +\frac{3}{2 \, \Big(1+(\frac{\omega_0}{\omega_{cm}})^2\Big)}\mbox{Im} \bigg[ \int\limits _0^{q_0}{d \, u} \, \frac{(q_0^2+u^2)}{q^3_0}\nonumber\\
&& \times\,\bigg(\cot{( u\, d)} -\frac{u^2}{q^2_0}(\cos{(2b\, u)} \, \cot{(u \, d)}+\sin{(2b \, u)} ) \bigg) \bigg]  \nonumber\\
&& + \, \frac{3 \, (1-\frac{\omega_0}{\omega_{cm}})^2}{1+(\frac{\omega_0}{\omega_{cm}})^2} \mbox{Im} \bigg[ \int _0^{q_0}{d \, u} \, \frac{u^2}{q^3_0}\nonumber\\
&& \times\,\bigg(\cot{( u\, d)} +\frac{u^2}{q^2_0}(\cos{(2b\, u)} \, \cot{(u \, d)}+\sin{(2b \, u)} ) \bigg) \bigg]  \bigg\}.
\end{eqnarray}
As a consistency check it can be easily seen that for the case where one of the plates goes to infinity ($d \rightarrow \infty$), Eq. (\ref{20}) is reproduced as expected.
\begin{figure}[htb!]
\includegraphics[width=18.0pc]{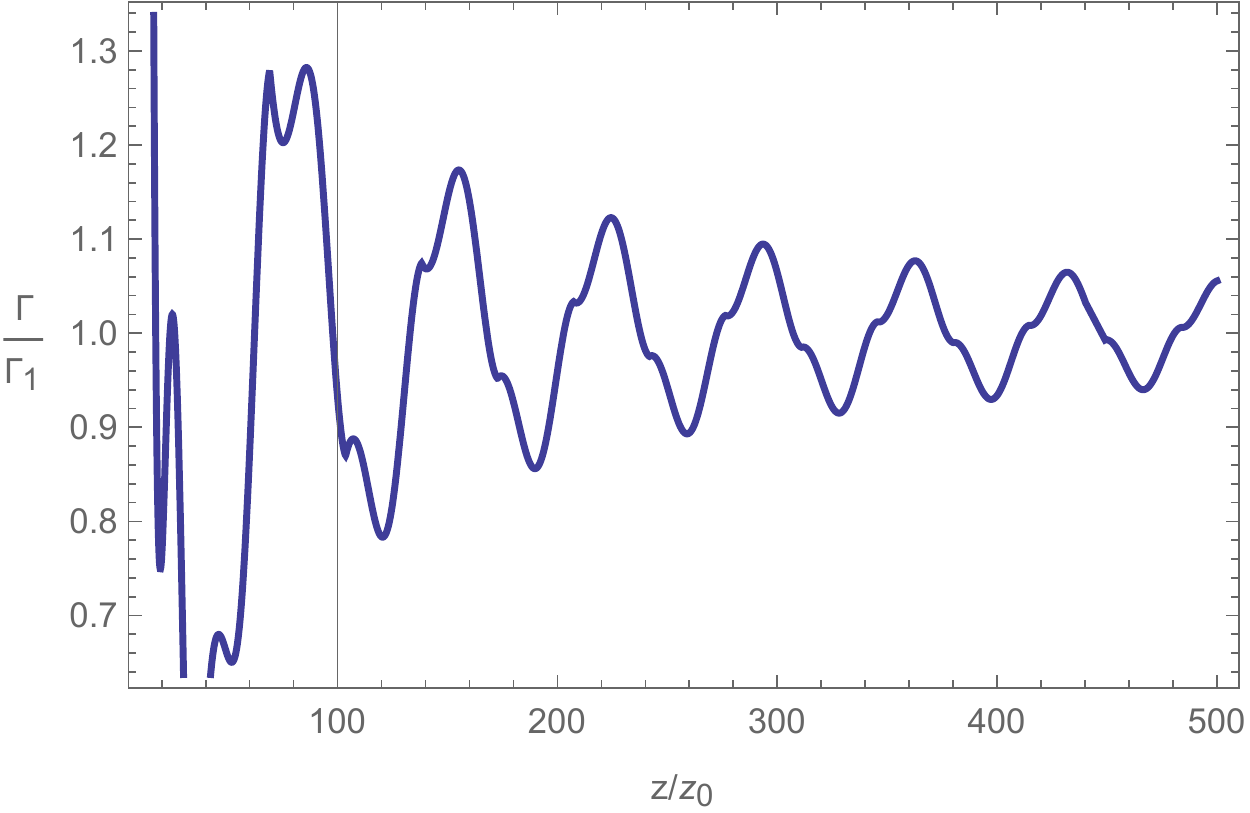}
\hspace{5mm}
\vspace{0.4cm}
\includegraphics[width=18.0pc]{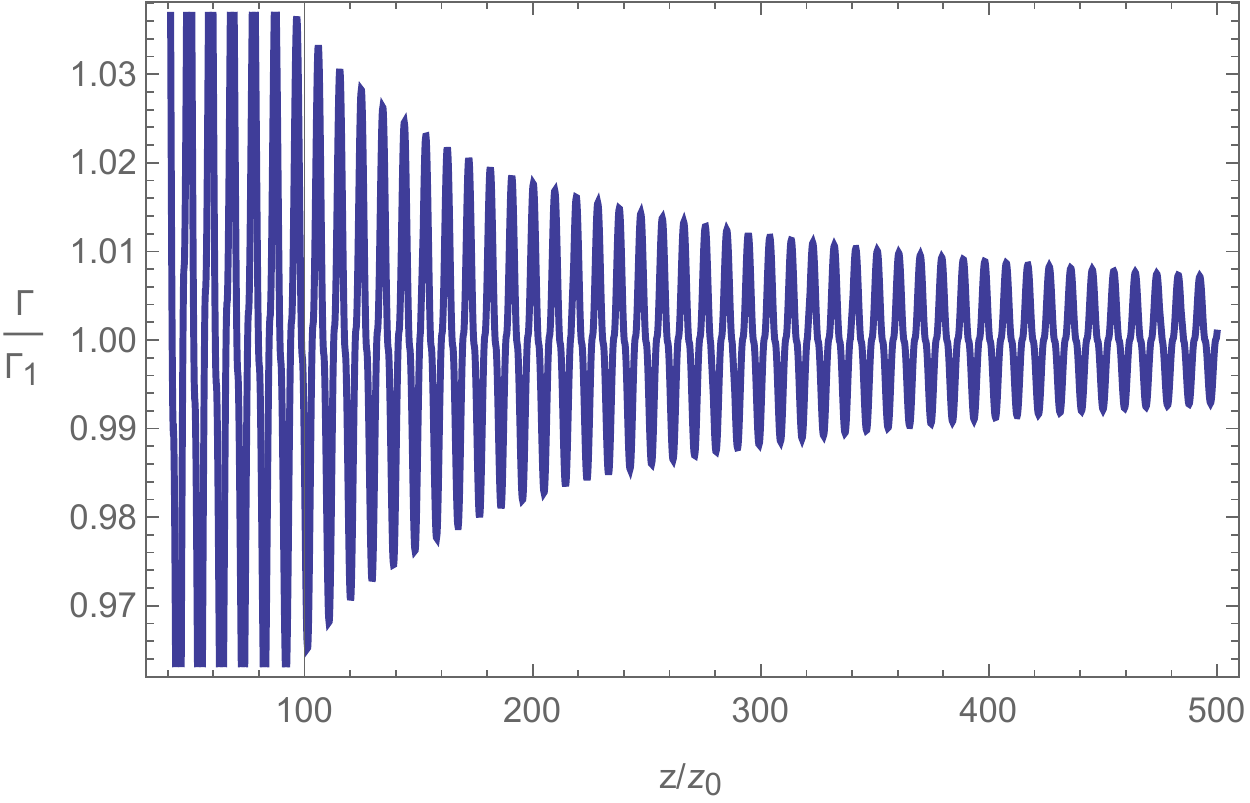}
\vspace{0.4cm}
\caption{(Color online) The scaled emission rate $\Gamma/\Gamma_1$ of a non-relativistic oscillating two-level atom located between two perfectly conducting plates ($b=d/2$) in terms of the scaled distance ($z/z_0= \, \omega_{cm} \, d/c$) for $\omega_{cm} = 1.1 \,\omega_0$ (left) and $\omega_{cm} = 3 \,\omega_0$ (right).} \label{fig4}
\end{figure}
In Fig. (4), the scaled emission rate $\Gamma/\Gamma_1$ of a non-relativistic oscillating two-level atom located between perfect parallel conductors ($b=d/2$) has been depicted in terms of the scaled distance ($z/z_0= \, \omega_{cm} \, d/c$) for $\omega_{cm} = 1.1 \, \omega_0$ and $\omega_{cm} = 3 \, \omega_0$.
\section{The decay rate of an oscillating atom }
In this section, we calculate the spontaneous decay rate of an initially excited two-level atom due to coupling to the vacuum field in vicinity of a perfect conductor. The atom oscillates in $z$ direction perpendicular to the conductor surface with frequency $\Omega$. The position of the center of mass of the atom is considered as
\begin{equation}\label{25}
 z=b+a \cos{(\Omega t)},
\end{equation}
with the velocity
\begin{equation}\label{26}
v=-v_{max}\sin{(\Omega t)},
\end{equation}
where $v_{max}=a\, \Omega$. We consider a fully quantum model and assumed that the atom is initially in its excited state denoted by $\left| e \right\rangle$ and the fluctuating field is in its vacuum state $\left| 0 \right\rangle$. In order to obtain a real description of atom-field interaction in the laboratory frame, we add the R\"{o}ngten term to the interaction Hamiltonian
\begin{eqnarray}\label{27}
&& H_{int}(t)=-\textbf{d} \cdot \Big(\textbf{E}(\textbf{r}(t))+\frac{\textbf{v}(t)}{c} \times \textbf{B}(\textbf{r}(t))\Big),
\end{eqnarray}
where the atomic dipole operator is defined by \cite{17}
\begin{eqnarray}\label{28}
&& \textbf{d}(0)=\left\langle g \right| \textbf{d}(0) \left| e \right\rangle \left( \left| g \right\rangle \left\langle e \right|+\left| e \right\rangle \left\langle g \right| \right).
\end{eqnarray}
The electric and magnetic field operators in the presence of an ideal conductor plate can be obtained as \cite{17}
\begin{eqnarray}\label{29}
&& \textbf{E}(\textbf{r},t)=\sum_{ \mathbf{k}, \lambda}-\sqrt{{2 \pi \hbar \omega_{\mathbf{k}}}} \, \textbf{f}(\textbf{r})_{ \mathbf{k}, \lambda}\, a_{ \mathbf{k}, \lambda}(t)+H.c\nonumber\\
&& \textbf{B}(\textbf{r},t)=\sum_{ \mathbf{k}, \lambda}i \, c \sqrt{\frac{2 \pi \hbar}{\omega_{\mathbf{k}} }}\Big({\nabla}\times \textbf{f}(\textbf{r})_{ \mathbf{k}, \lambda}\Big) \, a_{ \mathbf{k}, \lambda}(t)+H.c,
\end{eqnarray}
where $\mathbf{f}_{\mathbf{k},\lambda}$ is the spatial profile for two polarizations $\lambda= TE, TM$ given by
\begin{eqnarray}\label{30}
&& \textbf{f}_{ \mathbf{k},TE}=\sqrt{\frac{2}{V}}(\hat{\mathbf{k}}_{\parallel}\times \hat{\textbf{z}} \sin{k_z z})e^{i k_{\parallel} r},\nonumber\\
&& \textbf{f}_{ \mathbf{k} ,TM}=\sqrt{\frac{2}{V}}(\hat{\textbf{k}}_{\parallel}\frac{k_z}{k}\sin{k_z z}+i\hat{\textbf{z}} \frac{ k_{\parallel}}{k} \cos{k_z z})e^{i k_{\parallel} r},
\end{eqnarray}
In Eqs. (\ref{30}) $k_{\parallel}$ and $k_z$ refer to the parallel and perpendicular components of the wave vector $\mathbf{k}$ with respect to the conductor surface, respectively. The normalization or quantization volume is denoted by $V$. The decay rate of an initially excited atom can be obtained from the time-dependent perturbation theory as
\begin{eqnarray}\label{31}
&& \Gamma _{\left| e \right\rangle \rightarrow \left| g \right\rangle}= \frac{1}{T} \frac{1}{\hbar^2} \left|\int_0 ^T{\left\langle 1_{\mathbf{k}, \lambda},g\left|H_{int}(t)\right|0,e \right\rangle}dt\right|^2.
\end{eqnarray}
Now by inserting Eq. (\ref{27}) into Eq. (\ref{31}), and doing straightforward calculations, we find the decay rate of the oscillating atom given in the Appendix C. In the limiting case $a\rightarrow 0$, we recover the results reported in \cite{l3}
\begin{eqnarray}\label{32}
\Gamma_{\left| e \right\rangle \rightarrow \left| g \right\rangle} &=& \frac{4\left|d_{ge}\right|^2_{\bot} \omega_0^3}{3 \hbar \, c^3 }\left[1-3\left( \frac{\cos{(2 k_0 \, b)}}{(2 k_0 \, b)^2}-\frac{\sin{(2 k_0 \, b)}}{(2 k_0 \, b)^3}\right)\right]\nonumber\\
&& + \, \frac{4\left|d_{ge}\right|^2_{\parallel} \omega_0^3}{3 \hbar \, c^3 }\left[1-\frac{3}{2}\left(\frac{\sin{(2 k_0 \, b)}}{2 k_0 \, b} -\frac{\sin{(2 k_0 \, b)}}{(2 k_0 \, b)^3}+\frac{\cos{(2 k_0 \, b)}}{(2 k_0 \, b)^2}\right)\right].\nonumber\\
\end{eqnarray}
 \begin{figure}[htb!]
\centering
\resizebox{0.5\textwidth}{!}{%
  \includegraphics{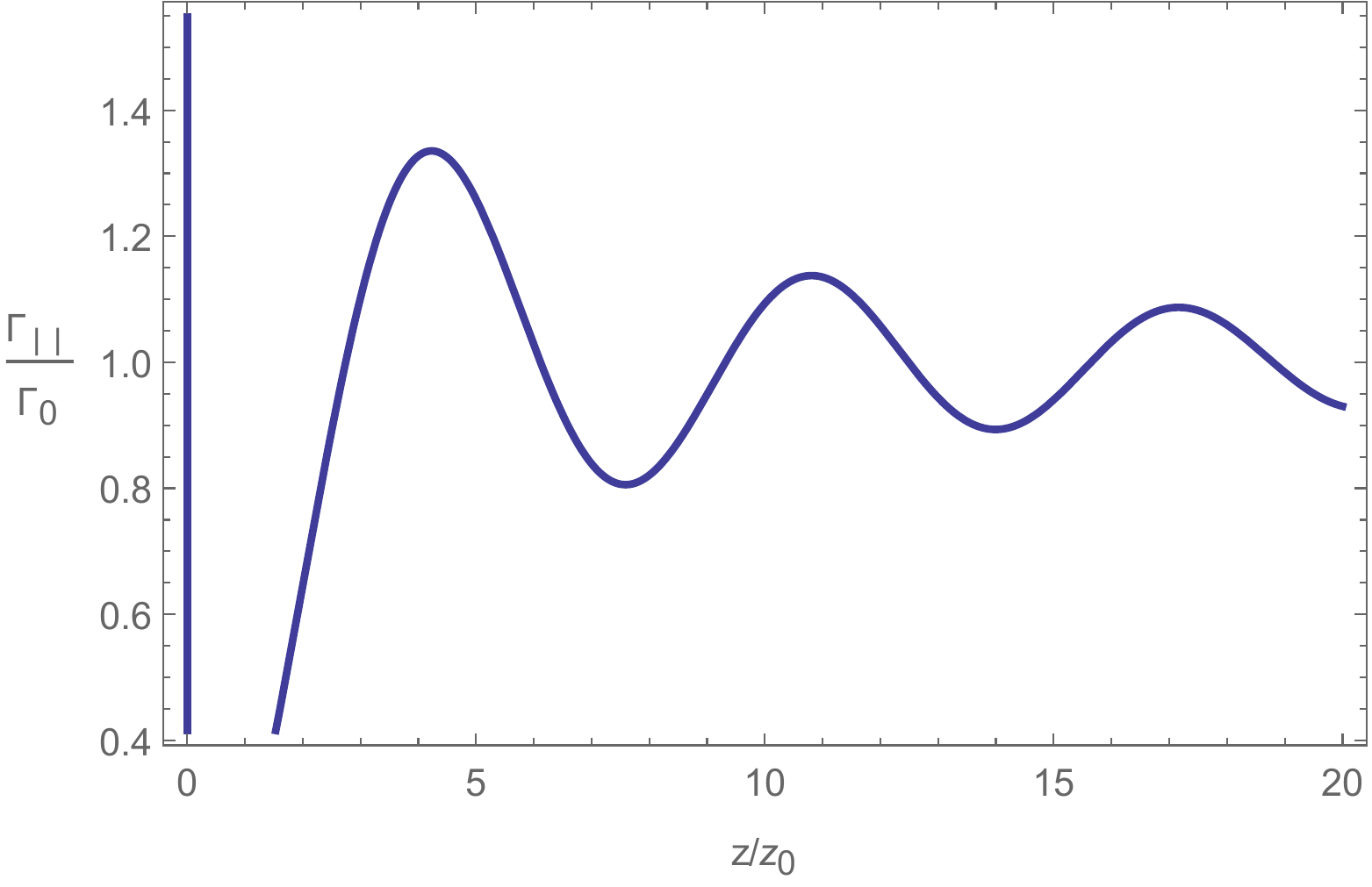}
}
\vspace{0.8cm}
\caption{Dimensionless decay rate of an initially excited oscillating atom between conducting plates in terms of
the dimensionless variable $z/z_0=2 \, k_0 \, b$ for $v_{max}=343\, m/s$ and parallel polarization}
\end{figure}
where $k_0= \omega_0/c$ ($\omega_0$ is the characteristic frequency of the atom). Also in the limits $a=0,\,b\rightarrow \infty$, the spontaneous emission rate tends to the vacuum spontaneous decay rate $ \Gamma_0= 4\left|\left\langle e\left| \mathbf{d}(0)\right|g\right\rangle\right|^2 {\omega_0}^3/3 \hbar c^3$ as expected.

In Figs. (5) and (6) the dimensionless decay rate for parallel and perpendicular polarizations of an initially excited oscillating atom near a conducting half-space is depicted in terms of the dimensionless variable $z/z_0=(2 \, k_0 \, b)$.
 \begin{figure}[htb!]
\centering
\resizebox{0.5\textwidth}{!}{%
  \includegraphics{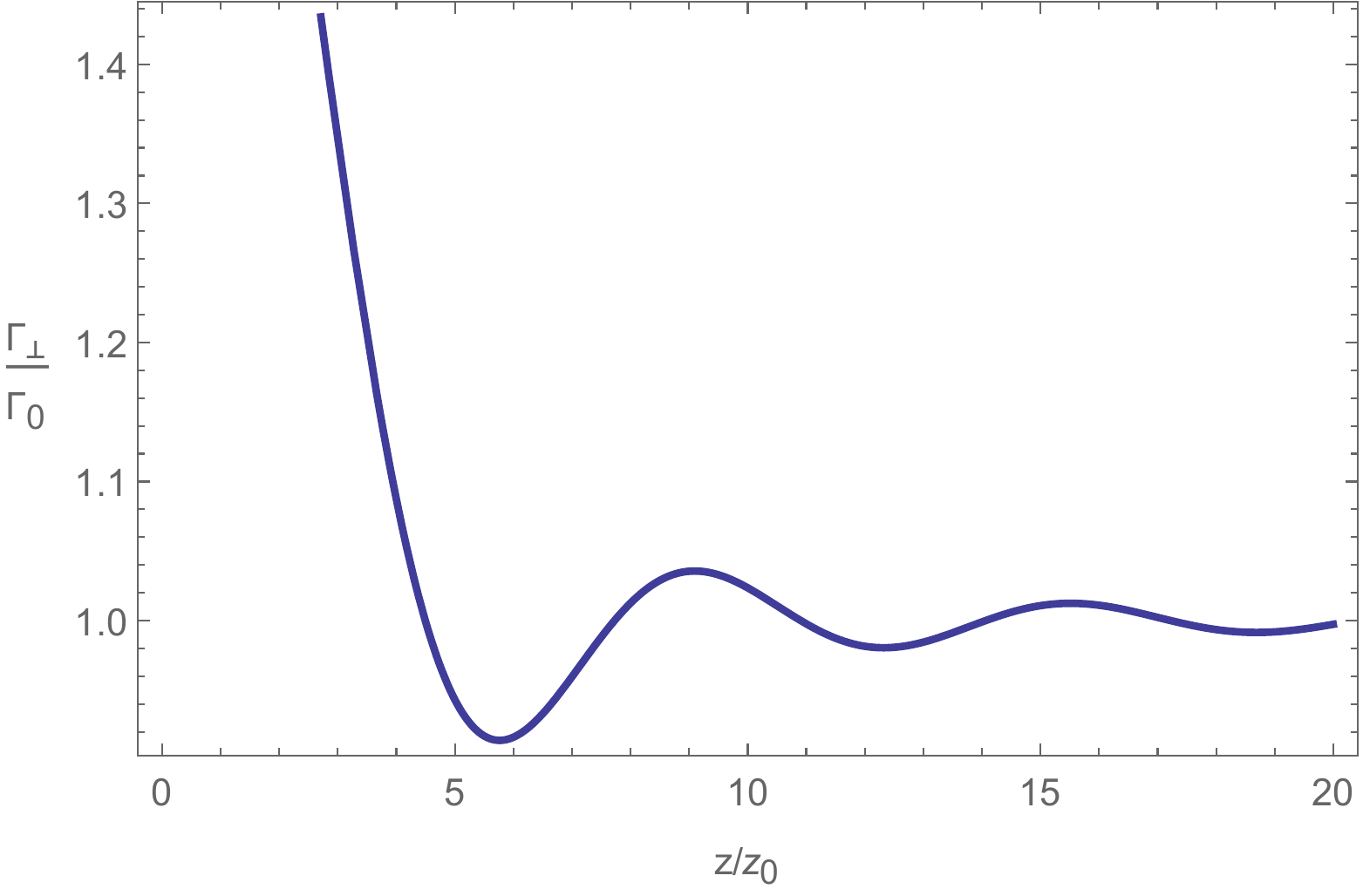}
}\vspace{.8cm}
\caption{Dimensionless decay rate of an initially excited oscillating atom between conducting plates in terms of
the dimensionless variable $z/z_0=2 \, k_0 \, b$ for $v_{max}=343\, m/s$ and perpendicular polarization}
\end{figure}
\section{conclusions}

The corrections to the photon emission rate of an oscillating two-level atom in the presence of electromagnetic quantum vacuum was investigated for two geometries: (i) Atom was trapped in vicinity of a perfect conductor (ii) Atom was trapped between two perfect conductors. The presence of conductors caused a damped oscillatory behaviour of photon emission rate around the no boundary result $\Gamma_1$ with maximum and minimum values appearing at regular distances from the conductors. The spontaneous decay rate of an initially excited atom oscillating in the vicinity of a perfect conductor was investigated. The decay rates had a damped oscillatory behaviour around the no boundary result $\Gamma_0$ at regular distances from the conductor.
\appendix
\section{}

Using the Fourier transform in $x$-$y$ directions, we define the dimensionally reduced dyadic tensor \cite{16}
\begin{eqnarray}
&&  \textbf{G}(\textbf{r},\textbf{r}',\omega)=\int{\frac{d^2 \mathbf{q}_{\parallel}}{(2\pi)^2}} e^{i \mathbf{q}_{\parallel}.(\mathbf{r}_{\parallel}-\mathbf{r}'_{\parallel})} \textbf{g}(z,z',\omega, \mathbf{q}_{\parallel}),
\end{eqnarray}
where $\mathbf{q}_{\parallel}=(q_x,q_y,0)$ and $\mathbf{r}_{\parallel}=(x,y,0)$. The diagonal components of $\textbf{g}$ matrix for conducting plate boundary conditions $(z,z' > 0)$ are given by \cite{16}
\begin{eqnarray}
g_{xx}(z(t),z(t'),\omega, \mathbf{q}_{\parallel}) &=& 2\pi\,i \frac{\eta q}{q^2} \bigg[e^{i \, \eta q |z-z'|} - e^{i \, \eta q (z+z')} \bigg], \nonumber\\
g_{yy}(z(t),z(t'),\omega, \mathbf{q}_{\parallel}) &=& 2\pi\,i \frac{1}{\eta q} \bigg[e^{i \, \eta q |z-z'|} - e^{i \, \eta q (z+z')} \bigg],\nonumber\\
g_{zz}(z(t),z(t'),\omega, \mathbf{q}_{\parallel}) &=& 2\pi\,i \frac{q_{\parallel}^2}{(\eta q )q^2} \bigg[e^{i \, \eta q |z-z'|} + e^{i \, \eta q (z+z')} \bigg]-\frac{4\pi }{q^2}\delta(z(t)-z(t')),\nonumber\\
\end{eqnarray}
where $\eta=\sqrt{1-\frac{q_{\parallel}^2}{q^2}}$. By inserting the coordinates given in Eq. (\ref{17}) into (A2), the components of the reduced Green's tensor up to the second order in $\alpha=q a$, are
\begin{eqnarray}
  g_{xx}(z(t),z(t'),\omega, \mathbf{q}_{\parallel}) &\approx & 2\pi\,i \frac{\eta q}{q^2} \bigg[(1-e^{2i\,b\sqrt{q^2-q_{\parallel}^2}})-i(\eta \alpha)(n e^{2i\,b\sqrt{q^2-q_{\parallel}^2}}-m)\nonumber\\
  && +\,\frac{1}{2}(\eta \alpha)^2(n^2 e^{2i b \sqrt{q^2-q_{\parallel}^2}}-m^2)\bigg],\nonumber\\
\end{eqnarray}
\begin{eqnarray}
 g_{yy}(z(t),z(t'),\omega, \mathbf{q}_{\parallel}) &\approx & 2\pi\,i \frac{1}{\eta q} \bigg[(1-e^{2i\,b\sqrt{q^2-q_{\parallel}^2}})-i(\eta \alpha)(n e^{2i\,b\sqrt{q^2-q_{\parallel}^2}}-m)\nonumber\\
 && +\,\frac{1}{2}(\eta \alpha)^2(n^2 e^{2i b \sqrt{q^2-q_{\parallel}^2}}-m^2)\bigg], \nonumber\\
\end{eqnarray}
and
\begin{eqnarray}
g_{zz}(z(t),z(t'),\omega, \mathbf{q}_{\parallel}) &\approx & 2\pi\,i \frac{q_{\parallel}^2}{\eta q^3} \bigg[(1+e^{2i\,b\sqrt{q^2-q_{\parallel}^2}})+i(\eta \alpha)(n e^{2i\,b\sqrt{q^2-q_{\parallel}^2}}+m)\nonumber\\
&& -\,\frac{1}{2}(\eta \alpha)^2(n^2 e^{2i b \sqrt{q^2-q_{\parallel}^2}}+m^2)\bigg]-\frac{4\pi }{q^2}\delta(z(t)-z(t')),\nonumber\\
\end{eqnarray}
where $m =\cos{(\omega_{cm}t)}-\cos{(\omega_{cm}t')}$ and $n=\cos{(\omega_{cm}t)}+\cos{(\omega_{cm}t')}$.

By inserting Eqs. (A3), (A4) and (A5) into Eq. (A1), and using Eqs. (\ref{13}), (\ref{14}) and (\ref{15}), we will find in large-time limit
\begin{eqnarray}
\Gamma^{EB} &=& 2\frac{\left| \left\langle e\left| \mathbf{d}(0)\right|g\right\rangle \right|^2}{3{\hbar}} \frac{v_{max}^2}{c^2} \, (\frac{\omega_0-\omega_{cm}}{\omega_{cm}})  \mbox{Re} \, \int_0^{q_0} du\, u^2,\nonumber\\
\Gamma^{EE} &=&  \frac{\left| \left\langle e\left| \mathbf{d}(0)\right|g\right\rangle \right|^2}{3 {\hbar}} \,\frac{v_{max}^2}{c^2} \frac{(\omega_{cm}-\omega_0)^2}{\omega_{cm}^2}   \bigg(\frac{q_0^3}{3}+ \mbox{Re} \, { \int_0^{q_0} {d u}\frac{u^4}{q_0^2}e^{2i u b}} \bigg),\nonumber\\
\Gamma^{BB} &=&   \frac{\left| \left\langle e\left| \mathbf{d}(0)\right|g\right\rangle \right|^2}{6 {\hbar}} \frac{v^2_{max}}{c^2}  \mbox{Re} \bigg[ \int _0^{q_0}{d \, u} \, (q_0^2+u^2) \bigg(1-e^{2i u \, b} \frac{u^2}{q^2_0} \bigg) \bigg],
\end{eqnarray}
where $q^2=q_z^2+q_{\parallel}^2$, $q_0=(\omega_{cm}-\omega_0)/c$, $v_{max}= \omega_{cm}a$, and $u=\sqrt{q_0^2-q_{\parallel}^2}$.
\section{}

The diagonal components of $\mathbf{g}$ in the presence of two parallel conducting plates for $0 < z, z' < d$ are given by \cite{16}
\begin{eqnarray}
g_{xx}(z(t),z(t'),\omega, \mathbf{q}_{\parallel}) &=& 2\pi\,i \frac{\eta q}{q^2} \bigg\{e^{-i \, \eta q |z-z'|} +\frac{1}{(1-e^{-2\,i \eta q \,d})} \bigg[ -e^{-i \, \eta q (z+z')}\nonumber\\
&& -\,e^{-2\,i \eta q \,d} e^{i \eta q (z+z')} + 2\, e^{-2\,i \eta q \,d} \cos{(\eta \, q(z-z'))}\bigg] \bigg\}\nonumber\\
&& -\,\frac{4\pi}{q^2} \delta(z(t)-z(t')),\nonumber\\
g_{yy}(z(t),z(t'),\omega, \mathbf{q}_{\parallel}) &=& 2\pi\,i \frac{1}{\eta q} \bigg\{e^{-i \, \eta q |z-z'|} +\frac{1}{(1-e^{-2\,i \eta q \,d})} \bigg[ -e^{-i \, \eta q (z+z')}\nonumber\\
&& -\,e^{-2\,i \eta q \,d} e^{i \eta q (z+z')}+ 2\, e^{-2\,i \eta q \,d} \cos{(\eta \, q(z-z'))} \bigg] \bigg\},\nonumber\\
g_{zz}(z(t),z(t'),\omega, \mathbf{q}_{\parallel}) &=& 2\pi\,i \frac{q_{\parallel}^2}{(\eta q )q^2} \bigg\{e^{-i \, \eta q |z-z'|} +\frac{1}{(1-e^{-2\,i \eta q \,d})} \bigg[ e^{-i \, \eta q (z+z')}\nonumber\\
&& +\,e^{-2\,i \eta q \,d} e^{i \eta q (z+z')}+ 2\, e^{-2\,i \eta q \,d} \cos{(\eta \, q(z-z'))} \bigg] \bigg\}\nonumber\\
&& -\,\frac{4\pi}{q^2} \delta(z(t)-z(t')).
\end{eqnarray}
By inserting Eq. (\ref{17}) into (B1), and expanding result up to the second order in $\alpha=q a$, we find
\begin{eqnarray}
g_{xx}(z(t),z(t'),\omega, \mathbf{q}_{\parallel}) &\approx & 2\pi\,i \frac{\eta q}{ q^2}\frac{e^{-2i\,b\sqrt{q^2-q_{\parallel}^2}}}{e^{2i\,d\sqrt{q^2-q_{\parallel}^2}}-1} \bigg[(-1+e^{2i\,b\sqrt{q^2-q_{\parallel}^2}})\nonumber\\
&\times & (-e^{2i\,b\sqrt{q^2-q_{\parallel}^2}}+e^{2i\,d\sqrt{q^2-q_{\parallel}^2}}) + i(\eta \alpha) \bigg(-n (e^{4i\,b\sqrt{q^2-q_{\parallel}^2}}\nonumber\\
&-& \,e^{2i\,d\sqrt{q^2-q_{\parallel}^2}})+me^{2i\,b\sqrt{q^2-q_{\parallel}^2}}(e^{2i\,d\sqrt{q^2-q_{\parallel}^2}}-1) \bigg)\nonumber\\
&+& \, \frac{1}{2}(\eta \alpha)^2 \Big(n^2 (e^{4i b \sqrt{q^2-q_{\parallel}^2}}+e^{2i\,d\sqrt{q^2-q_{\parallel}^2}})\nonumber\\
&-& \,m^2(e^{2i\,b\sqrt{q^2-q_{\parallel}^2}}+e^{2i\,b\sqrt{q^2-q_{\parallel}^2}}e^{2i\,d\sqrt{q^2-q_{\parallel}^2}}) \Big)\bigg]\nonumber\\
&-& \,\frac{4\pi}{q^2} \delta(z(t)-z(t')) \nonumber\\
\end{eqnarray}
\begin{eqnarray}
g_{yy}(z(t),z(t'),\omega, \mathbf{q}_{\parallel}) &\approx & 2\pi\,i \frac{1}{\eta q}\frac{e^{-2i\,b\sqrt{q^2-q_{\parallel}^2}}}{e^{2i\,d\sqrt{q^2-q_{\parallel}^2}}-1} \bigg[(-1+e^{2i\,b\sqrt{q^2-q_{\parallel}^2}})\nonumber\\
&\times & (-e^{2i\,b\sqrt{q^2-q_{\parallel}^2}}+e^{2i\,d\sqrt{q^2-q_{\parallel}^2}})+ i(\eta \alpha) \bigg(-n (e^{4i\,b\sqrt{q^2-q_{\parallel}^2}}\nonumber\\
&-& \,e^{2i\,d\sqrt{q^2-q_{\parallel}^2}})+me^{2i\,b\sqrt{q^2-q_{\parallel}^2}}(e^{2i\,d\sqrt{q^2-q_{\parallel}^2}}-1) \bigg)\nonumber\\
&+& \,\frac{1}{2}(\eta \alpha)^2 \Big(n^2 (e^{4i b \sqrt{q^2-q_{\parallel}^2}}+e^{2i\,d\sqrt{q^2-q_{\parallel}^2}})\nonumber\\
&-& \,m^2(e^{2i\,b\sqrt{q^2-q_{\parallel}^2}}+e^{2i\,b\sqrt{q^2-q_{\parallel}^2}}e^{2i\,d\sqrt{q^2-q_{\parallel}^2}}) \Big)\bigg] \nonumber\\
\end{eqnarray}
\begin{eqnarray}
g_{zz}(z(t),z(t'),\omega, \mathbf{q}_{\parallel}) &\approx & 2\pi\,i \frac{q_{\parallel}^2}{\eta q\, (q^2)} \frac{e^{-2i\,b\sqrt{q^2-q_{\parallel}^2}}}{e^{2i\,d\sqrt{q^2-q_{\parallel}^2}}-1} \bigg[(e^{2i\,b\sqrt{q^2-q_{\parallel}^2}}+1)\nonumber\\
&\times &\,(e^{2i\,b\sqrt{q^2-q_{\parallel}^2}}+e^{2i\,d\sqrt{q^2-q_{\parallel}^2}})+ \,i(\eta \alpha) \bigg(n (e^{4i\,b\sqrt{q^2-q_{\parallel}^2}}\nonumber\\
&-& \,e^{2i\,d\sqrt{q^2-q_{\parallel}^2}})+me^{2i\,b\sqrt{q^2-q_{\parallel}^2}}(e^{2i\,d\sqrt{q^2-q_{\parallel}^2}}-1) \bigg)\nonumber\\
&-& \,\frac{1}{2}(\eta \alpha)^2 \Big(n^2 (e^{4i b \sqrt{q^2-q_{\parallel}^2}}+e^{2i\,d\sqrt{q^2-q_{\parallel}^2}})\nonumber\\
&+& \,m^2(e^{2i\,b\sqrt{q^2-q_{\parallel}^2}}+e^{2i\,b\sqrt{q^2-q_{\parallel}^2}}e^{2i\,d\sqrt{q^2-q_{\parallel}^2}}) \Big)\bigg]\nonumber\\
&-& \,\frac{4\pi}{q^2} \delta(z(t)-z(t')) \nonumber\\
\end{eqnarray}
Now by inserting Eqs. (B2), (B3) and (B4) into Eq. (A1), and using Eqs. (\ref{13}), (\ref{14}) and (\ref{15}), we find the following results for the geometry of parallel conductors
\begin{eqnarray}
\Gamma^{EB} &=& \frac{2\left| \left\langle e\left| \mathbf{d}(0)\right|g\right\rangle \right|^2}{3{\hbar}} \frac{v_{max}^2}{c^2} \, (\frac{\omega_0-\omega_{cm}}{\omega_{cm}})  \mbox{Re} \, \int_0^{q_0} d q_{\parallel}\, (q_0^2-q_{\parallel}^2),\nonumber\\
\end{eqnarray}
\begin{eqnarray}
\Gamma^{EE} &=& \frac{1}{T} \frac{\left| \left\langle e\left| \mathbf{d}(0)\right|g\right\rangle \right|^2}{3 \pi{\hbar}c^2}\int_{0}^{\infty}{d \omega} \, (\omega \, a)^2 \int_{0}^{T}{d t}\int_{0}^{T}{d t'} e^{-i(\omega+ \omega_0)(t-t')}\nonumber\\
&\times &\, \mbox{Re} \, \bigg\{ \int_0^{q}{d q_{\parallel}} \, q_{\parallel} \, (\frac{\sqrt{q^2-q_{\parallel}^2}}{2 \, q^2})\frac{e^{-2i \eta q b}}{e^{2i \eta q d}-1}\bigg[-((q^2-q^2_{\parallel})+q^2+q_{\parallel}^2)\nonumber\\
&\times &\, \bigg( \big(e^{2i \eta b q}+e^{2i \eta b q}e^{2i \eta d q} \big) \big(\cos{(\omega_{cm}t)}-\cos{(\omega_{cm}t')} \big)^2\bigg) \nonumber\\
&+& \,\bigg((-q_{\parallel}^2+q^2+(q^2-q^2_{\parallel})) \big(e^{4i \eta q b}+ e^{2i \eta q d} \big) \nonumber\\
&\times & \,\big(\cos{(\omega_{cm}t)}+\cos{(\omega_{cm}t')} \big)^2 \bigg) \bigg]\bigg\},
\end{eqnarray}
\begin{eqnarray}
\Gamma^{BB} &=& \frac{1}{T} \frac{\left| \left\langle e\left| \mathbf{d}(0)\right|g\right\rangle \right|^2}{6 \pi{\hbar}} \,\frac{v_{max}^2}{c^2} \  \mbox{Re} \, \bigg\{ \int_{0}^{\infty}{d \omega} \int_{0}^{T}{d t}\int_{0}^{T}{d t'} e^{-i(\omega+ \omega_0)(t-t')}\nonumber\\
&\times &\, \sin{(\omega_{cm}t)} \, \sin{(\omega_{cm}t')}\int_0^{q}{d q_{\parallel}}\frac{q_{\parallel}}{\sqrt{q^2-q_{\parallel}^2}} \frac{2q^2-q_{\parallel}^2}{q^2}\nonumber\\
&\times &\, \bigg[ (q^2_{\parallel}) \bigg( \Big( \frac{1+e^{-2ib \eta q}}{e^{2id \eta q}-1} \Big) \Big(e^{2id \eta q}+e^{2ib \eta q} \Big) \bigg)\nonumber\\
&& + (2q^2-q_{\parallel}^2)\bigg( \Big(\frac{1-e^{-2ib \eta q}}{e^{2id \eta q}-1} \Big) \Big(e^{2id \eta q}-e^{2ib \eta q} \Big) \bigg) \bigg] \bigg\},
\end{eqnarray}
which lead to Eqs. (\ref{23}) after straightforward simplifications. The integrals in  Eqs. (\ref{23}) can be calculated leading to the following results
\begin{eqnarray}
\Gamma^{EB} &=& \frac{1}{6} \Gamma_0 \frac{v_{max}^2}{c^2} \Big(\frac{\omega_{cm}}{\omega_0}-1\Big)^3 \Big(\frac{\omega_0}{\omega_{cm}}-1\Big),
\end{eqnarray}
\begin{eqnarray}
\Gamma^{EE} &=& \frac{1}{4} \Gamma_0 \frac{v_{max}^2}{c^2}(\frac{\omega_{cm}}{\omega_0}-1)^3(1-\frac{\omega_0}{\omega_{cm}})^2 \frac{1}{12\, b^5d^5q_0^5} \bigg[ 6\, b \, q_0 \,d \cos{(2b\,q_0)}\nonumber\\
&\times &\, \bigg( d^4(-3+2b^2q_0^2)+ b^4 \Big( -2d^2q_0^2 \Big[\Phi(e^{2id\,q_0},2,\frac{-b}{d})+\Phi(e^{2id\,q_0},2,\frac{b}{d}) \Big]\nonumber\\
&+& 3\, \Big[ \Phi(e^{2id\,q_0},4,\frac{-b}{d})+\Phi(e^{2id\,q_0},4,\frac{b}{d}) \Big] \Big) \bigg)\nonumber\\
&+& \,4b^5d^3q_0^3 \bigg( d\,q_0 \Big(d\,q_0+3\, \mbox{Arg}\, (1-e^{-2id\,q_0}) \Big)+
 3\, {\mbox{Li}}_2 (e^{-2id\,q_0}) \bigg)\nonumber\\
&-& \,3\sin{(2b\, q_0)}   \bigg( 6\,b^5d^2q_0^2 \Big[\Phi(e^{2id\,q_0},3,\frac{-b}{d})-\Phi(e^{2id\,q_0},3,\frac{b}{d}) \Big]\nonumber\\
&+& \,3b^5 \Big[ \Phi(e^{2id\,q_0},5,\frac{b}{d})-\Phi(e^{2id\,q_0},5,\frac{-b}{d}) \Big]\nonumber\\
&+& \,d^5 \Big( -3+6b^2q_0^2-2b^4q_0^4+2b^4q_0^4 \Big[
 {}_2 F_1 (1,\frac{-b}{d},1-\frac{b}{d},e^{2id\,q_0})\nonumber\\
&+& \,{}_2 F_1(1,\frac{b}{d},1+\frac{b}{d},e^{2id\,q_0}) \Big] \Big) \bigg) \bigg],
\end{eqnarray}
and
\begin{eqnarray}
\Gamma^{BB} &=& \frac{1}{8} \Gamma_0 \frac{v_{max}^2}{c^2} \Big( \frac{\omega_{cm}}{\omega_0}-1 \Big)^3 \frac{1}{4\, b^5d^5q_0^5} \bigg[ 6\, b \, q_0 \,d \cos{(2b\,q_0)} \bigg( d^4(1-b^2q_0^2)\nonumber\\
&+& \,b^4 \Big( d^2q_0^2 \Big[ \Phi(e^{2id\,q_0},2,\frac{-b}{d})+\Phi(e^{2id\,q_0},2,\frac{b}{d}) \Big]\nonumber\\
&-& \,\Big[ \Phi(e^{2id\,q_0},4,\frac{-b}{d})+\Phi(e^{2id\,q_0},4,\frac{b}{d}) \Big] \Big) \bigg) \nonumber\\
&+& \,\frac{16}{3} \, b^5d^3q^3_0 \bigg(d\,q_0 \Big(d\,q_0+3\, \mbox{Arg}\, (1-e^{-2id\,q_0}) \Big)+
3\, {\mbox{Li}}_2 (e^{-2id\,q_0}) \bigg) \nonumber\\
&+& \,\sin{(2b\, q_0)} \bigg( d^5(-3+7b^2q_0^2-4\,b^4\,q_0^4)+b^4 \Big( 7\,b\,d^2q_0^2 \Big[\Phi(e^{2id\,q_0},3,\frac{-b}{d})\nonumber\\
&-& \,\Phi(e^{2id\,q_0},3,\frac{b}{d}) \Big]+3b \Big[ \Phi(e^{2id\,q_0},5,\frac{b}{d})-\Phi(e^{2id\,q_0},5,\frac{-b}{d}) \Big]\nonumber\\
&+& \, 4\,d^5q_0^4 \Big[{}_2 F_1(1,\frac{-b}{d},1-\frac{b}{d},e^{2id\,q_0})+{}_2 F_1(1,\frac{b}{d},1+\frac{b}{d},e^{2id\,q_0}) \Big] \Big) \bigg) \bigg].\nonumber\\
\end{eqnarray}
\section{}
By inserting Eq. (\ref{27}) into (\ref{31}), we find
\begin{eqnarray}
\Gamma_{\left| e \right\rangle \rightarrow \left| g \right\rangle} &=& \sum_{k, \lambda}\int^T_{0} {\int^T_{0} {dt}{dt'}}e^{i(\omega_k-\omega_0)(t-t')}\bigg\{2 \pi \hbar \omega_ \mathbf{k }\bigg[\textbf{d}_{ge} \cdot \textbf{f}_{ \mathbf{k}, \lambda}^{\ast}(\textbf{r}(t))\bigg]\nonumber\\
&\times &\, \bigg[\textbf{d}_{ge} \cdot \textbf{f}_{ \mathbf{k}, \lambda}(\textbf{r}(t')) \bigg]+\frac{2 \pi \hbar}{\omega_ \mathbf{k}}v_{max}^2 \bigg[\textbf{d}_{ge} \cdot \Big(\hat{\textbf{z}}\times [\textbf{k} \times \textbf{f}_{\textbf{k}, \lambda}^{\ast} (\textbf{r}(t)) ] \Big) \bigg]\nonumber\\
&\times &\, \bigg[\textbf{d}_{ge} \cdot \Big(\hat{\textbf{z}}\times[\textbf{k} \times \textbf{f}_{ \mathbf{k}, \lambda}(\textbf{r}(t'))] \Big) \bigg] \sin{(\Omega t)}\sin{(\Omega t')}\nonumber\\
&+& \,{2 \pi \hbar} v_{max} \bigg[\textbf{d}_{ge} \cdot \Big(\hat{\textbf{z}}\times[\textbf{k} \times \textbf{f}_{ \mathbf{k}, \lambda}^{\ast}(\textbf{r}(t))] \Big) \bigg] \bigg[\textbf{d}_{ge} \cdot \textbf{f}_{ \mathbf{k}, \lambda}(\textbf{r}(t')) \bigg]\sin{(\Omega t)}\nonumber\\
&+& \,{2 \pi \hbar} v_{max} \bigg[\textbf{d}_{ge} \cdot \Big(\hat{\textbf{z}}\times[\textbf{k} \times \textbf{f}_{ \mathbf{k}, \lambda}(\textbf{r}(t'))] \Big) \bigg] \bigg[\textbf{d}_{ge} \cdot \textbf{f}_{ \mathbf{k}, \lambda}^{\ast}(\textbf{r}(t)) \bigg]\sin{(\Omega t')} \bigg\},\nonumber\\
\end{eqnarray}
where $\omega_0$ is the internal frequency of the atom and $\textbf{d}_{ge}=\left\langle g \right| \textbf{d} \left| e \right\rangle $. By inserting Eq. (\ref{30}) into Eq. (C1) and expanding the result up to the second order in $k_z  a\ll 1$, we obtain
\begin{eqnarray}
\Gamma ^ {EE} _{\left| e \right\rangle \rightarrow \left| g \right\rangle} &=& \sum_{\mathbf{k}} \frac{ 4 \pi \omega_{\mathbf{k}}}{V}\frac{1}{T \hbar } \int_{0}^{T}{d t}\int_{0}^{T}{d t'} e^{i(\omega_{\mathbf{k}}-\omega_0)(t-t')}\bigg\{ |d_{ge}|^2_{\bot} \bigg[ \frac{k_{\parallel}^2}{k^2}(1+\cos{2 k_z b})\nonumber\\
&-& \,\frac{k_{\parallel}^2}{k^2}(\alpha' \, n' \sin{(2 k_z b)})+ \frac{k_{\parallel}^2}{k^2}(\frac{-(\alpha' m') ^2}{2})+ \frac{k_{\parallel}^2}{k^2} \cos{(2k_z b)} (\frac{-(\alpha' n')^2}{2})\bigg]\nonumber\\
&+& \,\frac{|d_{ge}|^2_{\parallel}}{2}\bigg[(1+\frac{k_{z}^2}{k^2})(1-\cos{(2k_z b)})+(1+\frac{k_{z}^2}{k^2})(\alpha' \, n' \sin{(2 k_z b)})\nonumber\\
&+& \,(1+\frac{k_{z}^2}{k^2})(\frac{-(\alpha' m') ^2}{2})-(1+\frac{k_{z}^2}{k^2})\cos{(2k_z b)}(\frac{-(\alpha' n')^2}{2})  \bigg] \bigg\}
\end{eqnarray}
\begin{eqnarray}
\Gamma ^ {BB} _{\left| e \right\rangle \rightarrow \left| g \right\rangle} &=& \sum_{\mathbf{k}} \frac{2 \pi }{\omega_{\mathbf{k}} \, V}v_{max}^2\frac{1}{T \hbar } \int_{0}^{T}{d t}\int_{0}^{T}{d t'} e^{i(\omega_{\mathbf{k}}-\omega_0)(t-t')} \sin{(\Omega t)} \, \sin{(\Omega t')} \nonumber\\
&& \times \bigg\{ |d_{ge}|^2_{\parallel}\bigg[ \frac{k_{\parallel}^4+k_z^4+k^2k_z^2}{k^2}+\frac{\cos{2 k_z b}}{k^2}(k_{\parallel}^4-k_z^4-k^2k_z^2)  \bigg] \bigg\},
\end{eqnarray}
\begin{eqnarray}
\Gamma ^ {EB} _{\left| e \right\rangle \rightarrow \left| g \right\rangle} &=& \sum_{\mathbf{k}} \frac{2 \pi}{ V}\frac{1}{T \hbar }v_{max} \int_{0}^{T}{d t}\int_{0}^{T}{d t'} e^{i(\omega_{\mathbf{k}}-\omega_0)(t-t')} \,  |d_{ge}|^2_{\parallel}\nonumber\\
&& \times \bigg\{ \bigg[\Big(\sin{(\Omega t')}-\sin{(\Omega t)} \Big)(\alpha' n')\nonumber\\
&&\hspace{1cm}+\Big(\sin{(\Omega t')}+\sin{(\Omega t)} \Big)(\alpha' m')\bigg]\frac{i k_{\parallel}^2 \, k_z}{k^2} \cos{(2k_z b)} \bigg\},
\end{eqnarray}
where
\begin{equation}
\alpha'=k_z a \qquad n'=\cos{(\Omega t)}+\cos{(\Omega t')} \qquad m'=\cos{(\Omega t)}-\cos{(\Omega t')}.
\end{equation}
Now integrals over time variables $t$, $t'$ can be calculated similar to (A7).  The summation over the modes can be approximated by integrals according to $\sum_ \mathbf{k} \rightarrow  \frac{V}{(2\pi)^3}\int{d^3 \mathbf{k}}$, and by changing the integration variable as $u=\sqrt{k^2-k_{\parallel}^2}$, we obtain
\begin{eqnarray}
\Gamma_{\left| e \right\rangle \rightarrow \left| g \right\rangle} &=& \Gamma^{EB}_{\left| e \right\rangle \rightarrow \left| g \right\rangle}+\Gamma^{EE}_{\left| e \right\rangle \rightarrow \left| g \right\rangle}+\Gamma^{BB} _{\left| e \right\rangle \rightarrow \left| g \right\rangle}\nonumber\\
&=& \int_0^{\omega_0/c}{du}\bigg\{ \frac{2\left|d_{ge}\right|^2_{\bot}}{ \hbar }(\frac{\omega_0^2}{c^2}-u^2) \bigg(1+\cos{(2u \, b)} \bigg)\nonumber\\
&+& \,\frac{\left|d_{ge}\right|^2_{\parallel}}{\hbar }(\frac{\omega_0^2}{c^2}+u^2) \bigg(1-\cos{(2u \, b)} \bigg)\bigg\}\nonumber\\
&+& \,\sum_{i=1,2}\int_0^{q_i}{du}{ \frac{\left|d_{ge}\right|^2_{\bot}}{\hbar}\frac{a^2}{2 } \bigg(1-\cos{}(2u \, b) \bigg)u^2(q_i^2-u^2)}\nonumber\\
&-& \,\int_0^{\omega_0/c}{du}{ \frac{\left|d_{ge}\right|^2_{\bot}}{\hbar}{a^2} \bigg(1+\cos{(2u \, b)} \bigg)u^2(\frac{\omega_0^2}{c^2}-u^2)}\nonumber\\
&+& \,\sum_{i=1,2}\int_0^{q_i}{du}{ \frac{\left|d_{ge}\right|^2_{\parallel}}{ \hbar}\frac{a^2}{4 } \bigg(1+\cos{(2u \, b)} \bigg)u^2(q_i^2+u^2)}\nonumber\\
&-& \,\int_0^{\omega_0/c}{du}{ \frac{\left|d_{ge}\right|^2_{\parallel}}{ \hbar}\frac{a^2}{2 } \bigg(1-\cos{(2u \, b)} \bigg)u^2(\frac{\omega_0^2}{c^2}+u^2)}\nonumber\\
&+& \,\sum_{i=1,2}\int_0^{q_i}{du} \frac{\left|d_{ge}\right|^2_{\parallel}}{ \hbar}\frac{v_{max}^2}{4\, c^2}\frac{1}{q_i^2} \bigg( \Big((q_i^2-u^2)^2+u^4+q_i^2u^2 \Big)\nonumber\\
&+& \,\cos{(2u \, b)} \Big((q_i^2-u^2)^2-u^4-q_i^2u^2 \Big)  \bigg) \nonumber\\
&+& \,\int_0^{q_1}{du} \frac{ \left| d_{ge} \right|^2_{\parallel}}{ \hbar}\frac{v_{max}}{2  \,c}\frac{a}{q_1}u^2(q_1^2-u^2)\bigg(\cos{(2u \, b)}+1  \bigg)\nonumber\\
&-& \,\int_0^{q_2}{du} \frac{ \left| d_{ge} \right|^2_{\parallel}}{\hbar}\frac{v_{max}}{2  \,c}\frac{a}{q_2}u^2(q_2^2-u^2)\bigg(\cos{(2u \, b)}+1  \bigg),\nonumber\\
\end{eqnarray}
where $q_1= (\omega_0+\Omega)/c$ and $q_2= (\omega_0-\Omega)/c$. From Eq. (C6), we have
\begin{eqnarray}
\frac{\Gamma_{\bot}}{\Gamma_0} &=& \bigg[1-3\left(\frac{\cos{(x)}}{x^2}- \frac{\sin{(x)}}{x^3}\right)+\frac{v_{max}^2(\xi^2)}{5 \, c^2 x^5}\nonumber\\
&\times &\, \bigg( \frac{1}{4}\Delta^5\,x^5+15(\Delta)x(-3+\frac{1}{4}x^2\Delta^2)\cos{ \Big(\Delta x \Big)}\nonumber\\
&& \hspace{0.2cm} + \, 15 \Big(3-\frac{5}{4}x^2(\Delta)^2 \Big)\sin{ \Big(x\Delta \Big)}+\frac{1}{4}\sigma^5x^5\nonumber\\
&& \hspace{0.2cm}+15\sigma x \Big(-3+\frac{1}{4}x^2\sigma^2 \Big)\cos{\Big(\sigma x \Big)}+15 \Big(3-\frac{5}{4}x^2(\sigma)^2 \Big)\sin{ \Big( x\sigma \Big)}\nonumber\\
&& - \, \frac{1}{2}x^5+30x(-3+\frac{1}{4}x^2)\cos{(x)}-30(-3+\frac{5}{4}x^2)\sin{(x)} \bigg) \bigg]
\end{eqnarray}
where $\Omega=\frac{1}{\xi} \, \omega_0$, $z/z_0 = x=2 \, k_0 \, b$, $\sigma =1+\frac{1}{\xi}$ and $\Delta=1-\frac{1}{\xi}$. Also
\begin{eqnarray}
\frac{\Gamma_{\parallel}}{\Gamma_0} &=& \bigg\{1-\frac{3}{2}\left(\frac{\sin{(x)}}{x} -\frac{\sin{(x)}}{(x)^3}+\frac{\cos{(x)}}{(x)^2}\right)\nonumber\\
&& + \,\frac{6 v_{max}^2}{c^2 \, x^5} \bigg[\xi ^2 \bigg[ -\frac{1}{30}x^5+\frac{1}{2}  \bigg(\cos{(x)}(-3x+\frac{3}{4}x^3)+\sin{(x)} \Big(3-x^2 \Big) \bigg)\nonumber\\
&& + \frac{1}{60} \bigg((x \, \Delta)^5+ 15(x \, \Delta)\cos{(x \, \Delta)} \Big(-3+\frac{3}{4}(x \, \Delta)^2 \Big)\nonumber\\
&& + \, 15 \Big(3-\frac{7}{4}(x \, \Delta)^2+\frac{1}{4}(x \, \Delta)^4 \Big) \sin{(x \, \Delta)}+ (\sigma x)^5\nonumber\\
&& + \,15(\sigma x)\cos{(\sigma x)} \Big(-3+\frac{3}{4}(\sigma x)^2 \Big)\nonumber\\
&& + \, 15 \Big(3-\frac{7}{4}(\sigma x)^2+\frac{1}{4}(\sigma x)^4 \Big) \sin{(\sigma x)} \bigg) \bigg]\nonumber\\
&& + \, \frac{1}{ \Delta ^2} \Bigg(\frac{1}{30}(x \, \Delta)^5+\frac{1}{16}(x \, \Delta)^2 \Big[ -3(x \, \Delta)\cos{(x \, \Delta)}\nonumber\\
&& + \,\Big( 3-(x \, \Delta)^2\Big) \sin{(x \, \Delta)}\Big] \bigg)+ \frac{1}{\sigma ^2} \Bigg(\frac{1}{30}(\sigma x)^5\nonumber\\
&& + \,\frac{1}{16}(\sigma x)^2 \Big[ -3(\sigma x)\cos{(\sigma x)}+ \Big( 3-(\sigma x)^2\Big) \sin{(\sigma x)}\Big] \bigg)\nonumber\\
&& - \, \frac{\xi}{30 \Delta} \Bigg(\frac{1}{4}(\Delta x)^5 -15(\Delta x) \Big( -3+\frac{1}{4}(\Delta x)^2\Big)\cos{(\Delta x)}\nonumber\\
&& - \,15 \Big( 3-\frac{5}{4}(\Delta x)^2\Big) \sin{(\Delta x)} \bigg)+\frac{\xi}{30\sigma}\Bigg(\frac{1}{4}(\sigma x)^5\nonumber\\
&& - \,15(\sigma x) \Big( -3+\frac{1}{4}(\sigma x)^2\Big)\cos{(\sigma x)}\nonumber\\
&& - \,15 \Big( 3-\frac{5}{4}(\sigma x)^2\Big) \sin{(\sigma x)} \bigg) \bigg] \bigg\}.
\end{eqnarray}


\section*{References}
\begin{thebibliography}{10}
\bibitem{1} Casimir H, 1948 Proc. K. Ned. Akad. Wet. \textbf{51} 793
\bibitem{2} Casimir H and Polder D 1948 Phys. Rev. \textbf{73} 360
\bibitem{3} Lifshitz E 1956 Zh. Eksp. Teor. Fiz. \textbf{29} 94
\bibitem{c1} Spaarnay M J 1958 Physica 24, 751
\bibitem{c2} Sabisky E S and Anderson C H 1973 Phys. Rev. A \textbf{7} 790
\bibitem{c3} Lamoreaux S K 1997 Physical Review Letters. \textbf{78} 5
\bibitem{c4} Mohideen U and Roy A 1998 Phys. Rev. Lett. \textbf{81} 4549;
Roy A and Mohideen U 1999 ibid. \textbf{82} 4380
\bibitem{c5} Milton K A, Deraad L L, and Schwinger J 1978, Ann. Phys. (NY)
\textbf{115}, 388
\bibitem{c6} Milton K A 2001 The Casimir Effect: Physical Manifestation of
Zero-Point Energy (World Scientific: Singapore)
\bibitem{c7} Mostepanenko V M and Trunov N N 1997 The Casimir Effect and
its Applications (Clarendon Press: Oxford)
\bibitem{l1} Bethe H A 1947 Phys. Rev. \textbf{72} 339
\bibitem{l2} Lamb W E, Retherford R C 1947 Physical Review. \textbf{72} 241
\bibitem{l3} Matloob R 2000 Phys. Rev. A 62, 022113
\bibitem{l4} Mohammadi Z and Kheirandish F 2015 Phys. Rev. A \textbf{92} 062118
\bibitem{4} Kroll N M and Lamb W E 1949 Phys. Rev. \textbf{75}, 388
\bibitem{5} Welton T A 1948 Phys. Rev. \textbf{74} 1157
\bibitem{6} Power E A 1966 Am. J. Phys. \textbf{34} 516
\bibitem{7} Agarwal G S 1975 Phys.Rev.A \textbf{11} 230; ibid. \textbf{11}, 243; ibid. \textbf{11} 253; ibid. \textbf{12} 1475; ibid. \textbf{12} 1974; ibid. \textbf{12} 1987
\bibitem{8} Milonni P W 1994 The Quantum Vacuum (Academic Press: New York)
\bibitem{9} Vogel V and Welsch D G Quantum Optics (Wiley-VCH: Berlin)
\bibitem{r1} Golestanian R Phys. Rev. Lett. \textbf{95}, 230601
\bibitem{k} Kheirandish F and Jafari M 2012 Phys. Rev. A \textbf{86} 022503; Kheirandish F and Salimi S 2011 Phys. Rev. A \textbf{84} 062122;
            Kheirandish, Soltani M and Sarabadani J 2011 Annals of Physics, \textbf{326} 657-667; Kheirandish F, Soltani M and Sarabadani J 2010 Phys. Rev. A \textbf{81}, 052110
\bibitem{10} Hinds E 1994 in Cavity Quantum Electrodynamics, Supplement to Advances in Atomic, Molecular, and Optical Physics, edited by P. R. Berman (Academic Press: New York); Heinzen D J and Feld M S 1987 Phys. Rev. Lett. \textbf{59} 2623
\bibitem{ors} Orszag M 2016 Quantum Optics, Third Edition, (Springer: Switzerland)
\bibitem{opto} Bowen W P and Milburn G J 2016 Quantum optomechanics (Taylor and Francis Group: NY)
\bibitem{Aha-Bohm} Horsley S and Babiker M 2005 Phys. Rev. Lett. \textbf{95}, 010405
\bibitem{Wilkens} Wilkens M 1994 Phys. Rev. A\textbf{ 49} 570
\bibitem{Barton} Barton G and Calogeracos A 1999 in ”The Cashimer Effect 50 Years Later”, Ed. M. Bordag (World Scientific: Singapore)
\bibitem{Muller} Audretsch J and M\"{u}ller R 1994 Phys. Rev. A \textbf{50} 1755
\bibitem{Unruh} Unruh W G and Wald R M 1984 Phys. Rev. D \textbf{29}, 1047
\bibitem{11} Souza R M e, Impens F, Neto P A M 2018 Phys. Rev. A \textbf{97} 032514
\bibitem{14} Landau L D and Lifshitz E M 1980 Statistical Physics (Pergamon: Oxford)
\bibitem{15} Matloob R 1999 Phys. Rev. A \textbf{60} 3421
\bibitem{16} Parashar P 2011, Geometrical investigations of the Casimir effect: Thickness and corrugations dependencies, Ph.D. thesis, (The University of Oklahoma: Norman)
\bibitem{17} Steck D A 2018 Quantum and Atom Optics, available online at http://steck.us/teaching (revision 0.12.3)
\end{thebibliography}
\end{document}